\documentstyle[amssymb,amstex,amscd,11pt]{article}

\def\bc{\begin{center}}
\def\ec{\end{center}}
\def\be{\begin{equation}}
\def\ee{\end{equation}}
\def\ben{\begin{enumerate}}
\def\een{\end{enumerate}}
\def\bfg{\begin{figure}}
\def\efg{\end{figure}}
\def\bq{\begin{quote}}
\def\eq{\end{quote}}
\def\bd{\begin{description}}
\def\ed{\end{description}}
\def\this{i.\ e.\ } 

\def\p{\partial}
\def\w{\wedge}

\def\dim{\operatorname{dim}}
\def\codim{\operatorname{codim}}

\def\det{\operatorname{det}}

\def\tr{\operatorname{tr}}

\newcommand{\CC}{{\Bbb C}}
\newcommand{\RR}{{\Bbb R}}

\newcommand{\ZZ}{{\Bbb Z}}
\newcommand{\QQ}{{\Bbb Q}}

\newcommand{\lan}{\langle}
\newcommand{\ran}{\rangle}

\newcommand{\ga}{\alpha}

\newcommand{\gc}{\gamma}
\newcommand{\gd}{\delta}
\newcommand{\gel}{\varepsilon}
\newcommand{\gf}{\varphi}

\newcommand{\gl}{\lambda}

\newcommand{\go}{\omega}
\newcommand{\gr}{\rho}
\newcommand{\gs}{\sigma}

\newcommand{\gz}{\zeta}

\newcommand{\gC}{\Gamma}
\newcommand{\gD}{\Delta}
\newcommand{\gL}{\Lambda}
\newcommand{\gO}{\Omega}
\newcommand{\gS}{\Sigma}
\newcommand{\cala}{{\cal A}}

\newcommand{\cali}{{\cal I}}
\newcommand{\calj}{{\cal J}}

\newcommand{\call}{{\cal L}}
\newcommand{\calm}{{\cal M}}

\title{QUANTUM COHOMOLOGY OF FLAG MANIFOLDS \\ AND TODA LATTICES}

\author{Alexander GIVENTAL \thanks{Supported by  Alfred P. Sloan Foundation}
\\ UC\, Berkeley
\and Bumsig KIM \\ UC\, Berkeley }

\date{ December 12, 1993}

\begin{document}

\maketitle

\begin{abstract}{ We discuss relations of Vafa's quantum cohomology
with Floer's homology theory, introduce equivariant quantum
cohomology, formulate some conjectures about its general
properties and, on the basis of these conjectures, compute quantum
cohomology algebras of the flag manifolds.  The answer turns out to coincide
with the algebra of regular functions on an invariant lagrangian variety of
a Toda lattice.}
\end{abstract}

\section{Introduction}

Quantum cohomology of compact complex Kahler manifolds was introduced
by C.Vafa [V] in connection with the theory of mirror manifolds.

By Vafa's definition, the quantum cohomology $QH^*(X)$ of a compact Kahler
manifold $X$ is a certain deformation of the cup-product multiplication in
the ordinary cohomology of $X$. Let $a,b,c$ be three cycles in $X$ representing
three given cohomology classes by Poincare duality. One defines the {\em
quantum
cup-product} $a*b$ by specifying its intersection indices with all $c$. Namely
\[ \lan a*b, c\ran = \sum _{\text{degree $d$ discrete holomorphic maps}:
                           \ (\CC P^1,0,1,\infty ) \to (X, a,b,c)} \pm q^d .\]
In other words, the intersection index takes in account rational parametrized
curves in $X$ with the three marked points --- images of $0$,$1$ and $\infty $
--- on the three cycles $a$, $b$ and $c$ respectively.

\bigskip
This definition needs some explanations.

\ben
\item First of all, a rational curve
contributes to the intersection index only if it is ``discrete'' which means,
by definition, that
\[ c(d) + \dim X = \codim a + \codim b + \codim c \]
where $c(d)$ is the first Chern class $c$ of (the tangent bundle to)
$X$ evaluated on the homology class $d$ of the curve,
$\dim X$ is the complex dimension of $X$, and {\em codim}  on the RHS stand for
degrees of the cohomology classes represented by $a,b,c$, also counted in
{\em complex} units (so that a real hypersurface has codimension $1/2$).
The meaning of the LHS is the dimension of the parameter space of such curves
predicted by the classical Riemann--Roch formula, while the RHS is the number
of constraints imposed at $0$,$1$ and $\infty $. Thus in the situation of
``general position'', when the Riemann--Roch prediction is correct (and under
some further transversality assumptions) the ``discrete'' curves can really be
treated as isolated intersections and contribute to $\lan a*b, c\ran $ by
$\pm q^d$ each.

\item Here ``$q^d$'' is, formally speaking, the homology class of the
rational curve
and therefore the intersection index as a whole is an element of a group ring
of the lattice $H_2(X,\ZZ )\cap H_{1,1}(X,\CC )$. The notation $q^d$ is
chosen simply to ``tame''
the group ring by means of coordinates on the lattice.
If we choose a basis of Kahler forms $\go _1,...,\go _k$ in
$H^2(X,\ZZ ) \cap H^{1,1}(X,\CC ) $ and express the homology class of a
rational curve $S$ by the string $d=(d_1,...,d_k)$ of its coordinates in the
dual basis (so that $d_i=\int _S \go _i \geq 0$) then the element $q^d$ of the
group ring can be identified with the monomial $q_1^{d_1}...q_k^{d_k}$ of the
formal variables $(q_1,...,q_k)$, and the intersection index $\lan a*b,c\ran $
becomes a formal series in $q$.

\item The constant term of this series counts
{\em constant} rational curves with the marked points in
the cycles $ a,b,c $, \this it counts ordinary intersection points.
The signs $\pm $ should be chosen in such a way that this term is the
ordinary triple intersection index $\lan a\cap b, c\ran $ of the cycles.

\item About
the higher degree terms (they are called ``instanton corrections'' to the
classical intersection index) we only tell here that their signs $\pm $ are
defined to be pluses only in the case when the cycles $a,b,c$ are complex
submanifolds in $X$ (while the general case will be briefly discussed in
$2.3$). In any way, the instanton corrections provide a $q$-deformation of
the classical triple intersection index.

\item The double intersection index
$\lan a, c \ran $ of any two cycles, by definition, coincides with the ordinary
non-degenerate Poincare pairing, and one can recover the quantum cup-product
$a*b$ from the triple pairings as an element of $H^*(X,\ZZ [[q]]) $.

\een

\bigskip
The above construction of the quantum cohomology ring is lacking of
many ingredients which could possibly make it mathematically rigorous, and
we will touch some mathematical aspects of the problem in the next section.
On the other hand, Vafa's construction is strongly supported by general
ideology of Conformal Topological Field Theory and provides mathematicians
with a bunch of interrelated conjectures. In particular, according to these
conjectures, the quantum cup-product

\begin{itemize}
\item can be defined rigorously;
\item is associative and skew-commutative;
\item is a $q$-deformation of the classical cup-product;
\item respects the usual grading in the cohomology provided that
      one assigns the following non-trivial degrees to the parameters
      of the deformation: $\deg q^d=c(d)$ (in complex units).
\end{itemize}

In this paper, we do not have any intention to justify these properties
mathematically. Instead, our objective is to compute the quantum cohomology
algebras of the classical flag manifolds {\em in the assumption that their
properties expected on the basis of Topological Field Theory are valid}.
Therefore the results obtained in this way, while ``physical theorems'',
have the status of mathematical conjectures, or better to say
{\em conditional} theorems contingent to the general conjectures about
quantum cohomology of Kahler manifolds. With this reservation in mind we
formulate below the results of our computation as {\em theorems}.

\bigskip
Let $F_{n+1}$ denote the manifold of complete flags
\[ \CC ^1 \subset ... \subset \CC ^n  \]
in $\CC ^{n+1}$. The cohomology algebra $H^*(F_{n+1})$ is known to be
canonically isomorphic to the quotient of the polynomial algebra
$\ZZ [u_0,...,u_n]$ in $n+1$ indeterminates by the ideal generated by
the elementary symmetric polynomials $\gs _1(u),...,\gs _{n+1}(u)$.
The generators $u_i$ are in fact the $1$-st Chern classes of the tautological
line bundles over the flag manifold with the fiber $\CC ^{i+1}/\CC ^i$. They
are constrained by $u_0+...+u_n=0$ and can be expressed through another basis
as $u_i=p_i-p_{i+1}$. The generators $(p_1,...,p_n)$ are $1$-st Chern classes
of the {\em determinant} line bundles with the fiber $\gL ^*\CC ^i$ over a
point $\CC ^1 \subset ... \subset \CC^n$ of the flag manifold. These
determinant line bundles are {\em non-negative} and the classes $p_i$ span
the edges of the (simplicial) Kahler cone in the $2$-nd cohomology of
$F_{n+1}$.
For a rational curve $S\subset F_{n+1}$ we define its {\em degree}
$d=(d_1,...,d_n)$ with respect the coordinates $p_i$ as
$d_i=\lan p_i, [S] \ran \geq 0$. Now the homology class of the curve is
represented by the monomial $q^d=q_1^{d_1}...q_n^{d_n}$.

In order to describe the quantum cohomology algebra $QH^*(F_{n+1})$ it suffices
therefore to exhibit the corresponding deformation of elementary symmetric
polynomials of $u_0,...,u_n$ by the parameters $q_1,...,q_n$. Notice that
while the degrees of $u_i$ are equal $1$, the degrees of all $q_i$ are equal
$2$
(since the $1$-st Chern class of the flag manifold is $c=2(p_1+...+p_n)$), and
the deformation should be homogeneous with respect to this grading.

\bigskip
Consider the diagonal matrix with $u_0,...,u_n$ on the diagonal. Then the
coefficients of its characteristic polynomial are elementary symmetric
functions of $u$.

Consider another $(n+1)\times (n+1)$ matrix, denoted $A_n$,
\[ A_n=\left[ \begin{array}{ccccc} u_0 & q_1 &  0  & ... &  0  \\
                                   -1  & u_1 & q_2 & ... &  0  \\
                                    0  & -1  & u_3 & ... &  0  \\
                                       &  .  &  .  &  .  &     \\
                                    0  & ... &  0  &  -1 & u_n
              \end{array} \right] \]
with $u_i$ on the diagonal, $q_i$ --- right above, and $-1$'s --- right under
the diagonal. Then the coefficients of its characteristic polynomial are the
deformations in question of the elementary symmetric functions:

\bigskip
{\bf Theorem 1.} {\em The quantum cohomology algebra $QH^*(F_{n+1})$ of the
flag manifold is canonically isomorphic to the quotient of the polynomial
algebra $\ZZ [u_0,...,u_n,q_1,...,q_n]$ by the ideal generated by
coefficients of the characteristic polynomial of the matrix $A_n$.}

\bigskip
Specialists on complete integrable systems will recognize in this answer
something very familiar: in fact the coefficients of $\det (A_n + \gl )$ are
conservation laws of a Toda lattice.

Namely, introduce ``configuration'' variables $(x_0,...,x_n)$ of $n+1$
consequtive unit masses on the line with $q_i=\exp(x_i-x_{i-1})$  in the role
of potential energy of neighbors. Then
\[ \frac{1}{2}\tr (A_n^2) = \frac{1}{2}\sum u_i^2 - \sum e^{x_i-x_{i-1}} \]
is the Hamiltonian of the classical Toda lattice (with incorrect sign of the
potential however), and $\tr (A_n^i), \ i=1,...,n+1$, is the complete set
of commuting first integrals.

\bigskip
{\bf Corollary.}{\em The quantum cohomology algebra of the flag manifold
$F_{n+1}$ is isomorphic to the algebra of functions on the common zero level
of the first integrals of the classical Toda lattice.}

\bigskip
Making comments on the theme ``How much surprising is the result?'' we should
say that one might not expect quantum cohomology of flag manifolds to have no
connections with other known objects attributed to flag manifolds. Moreover,
Topological Field Theory predicts  deep relations (see for instance [D],[W])
of moduli spaces of rational
curves in Kahler manifolds with hierarchies of integrable systems. Moreover,
Toda lattices have already occurred [CV] --- in a ``less surprising'' manner
---
in some dynamical problem related to quantum cohomology of projective
spaces. Nevertheless the authors should confess they did not foresee this
particular relation when started the computation, and they do not know now
how the answer can be predicted. However some partial explanations should be
given right away.

First of all, it can be viewed accidental that the relations in quantum
cohomology of flag manifolds Poisson-commute. What is not accidental at all is
that they Poisson-commute {\em modulo the relations themselves}. Indeed,
according to general theory (see 2.4) quantum cohomology algebra
of a Kahler manifold in some sense always is (or at least related to) the
algebra of functions on some lagrangian variety in the cotangent bundle of
some torus. The parameters $q_i$ of the quantum deformation are multiplicative
coordinates on the torus. In the case of $F_{n+1}$ the cotangent bundle
provided with the coordinates $q_1,...,q_n \neq 0, p_1,...,p_n$ (in above
notations) has the canonical symplectic form
\[ d p_1 \w \frac{dq_1}{q_1} + ... + d p_n \w \frac{dq_n}{q_n} ,\]
and the algebra $QH^*(F_{n+1},\CC )$ must be the algebra of regular functions
on some quasi - homogeneous lagrangian subvariety $L$. In view of the
group-theoretic nature of Toda lattices [R], our theorem leads to the
following geometrical description of $L$.

\bigskip
Let $G=SL_{n+1}(\CC )$, $N_+$ and $N_-$ be its strictly lower- and
upper-triangular subgroups. Make $N_+$ and $N_-$ act respectively by left
and right translations on the cotangent bundle $T^*G$ of the group and
consider the momentum map $J: T^*G \to Lie^*(N_+\times N_-)$ of the action.
The trace inner product $\tr A B$ on the matrix algebra identifies the dual
of the Lie algebra of $N_+\times N_-$ with the quotient of the space of all
square $(n+1)$-matrices by the subspace of all diagonal matrices. Pick the
value of the momentum map as specified by the matrix
\[ P = \left[ \begin{array}{ccccc} * & 1 & 0 & 0 & ... \\
                                     1 & * & 1 & 0 & ... \\
                                     0 & 1 & * & 1 & ... \\
                                       & . & . & . &     \\
                                    ...& 0 & 0 & 1 &  *
\end{array} \right] \]
($0$'s everywhere except $1$'s right above and under the diagonal) and
make the symplectic reduction on this level of the momentum map. The reduced
phase space
\[ M_P = T^*G//_P (N_+\times N_-) = J^{-1}(P) / (N_+\times N_-) \]
can be naturally identified with the cotangent bundle of the maximal torus in
$G$. Now, consider the cone $C\subset Lie\, G$ of all nilpotent traceless
matrices. The product
\[ C\times G \subset (Lie\, G) \times G \ =\  T^*G \]
is a bi-invariant involutive subvariety. Its symplectic reduction
\[ L = [J^{-1}(P) \cap (C\times G)]/(N_+\times N_-) \subset M_P \]
is in fact a lagrangian subvariety in the reduced phase space.

\bigskip
{\bf Corollary.} {\em The quantum cohomology algebra $QH^*(F_{n+1}, \CC )$
is isomorphic to the algebra of regular functions on the lagrangian variety
$L$.}

\bigskip
We should augment this corollary with an open question: {\em Why} the
quantum cohomology algebra of the flag manifold $G/B_-$ is isomorphic to
the algebra of regular functions on the lagrangian variety $L$?
We would expect that a natural answer to this question will come along with
a better understanding of the general mirror symmetry
phenomena (cf. [G3]).

\bigskip
The second argument that partially explains the theorem comes from its proof.
Our computation of quantum cohomology of flag manifolds is based in fact on
induction on $n$. It turns out however that the induction assumption that
quantum cohomology of $F_{m+1}$ with $m<n$ is known, is insufficient for our
purpose. What we really need is an {\em equivariant} version of quantum
cohomology of flag manifolds considered as homogeneous spaces of unitary
groups.
Similarly to ordinary equivariant cohomology of a $U$-space $X$, quantum
equivariant cohomology can be defined (with similar reservations) as a
skew-commutative associative algebra over the ring of characteristic classes
of the compact Lie group $U$.

In the case of $U=U_{n+1}$ (acting on the flag manifold $F_{n+1}$), we deal
with the algebra $\ZZ [c_1,...,c_{n+1}]$ of usual Chern classes, and the
ordinary equivariant cohomology of the flag manifold is known to coincide
with the polynomial algebra $\ZZ [u_0,...,u_n]$ of characteristic classes
of the maximal torus $T^{n+1}\subset U_{n+1}$ considered however as a module
over the subalgebra of Chern classes
\[ c_i=\gs _i(u_0,...,u_n),\ i=1,...,n+1 \]
--- elementary symmetric functions of $u$.

In the same manner as $H^*(F_{n+1})$ is obtained from the equivariant
cohomology $H^*_{U_{n+1}}(F_{n+1})$ by specialization $c_1=...=c_{n+1}=0$,
we deduce our theorem on quantum cohomology of flag manifolds from a more
general result describing their equivariant quantum cohomology.

\bigskip
{\bf Theorem 2.} {\em The equivariant quantum cohomology algebra
$QH^*_{U_{n+1}}(F_{n+1})$ is canonically isomorphic to the quotient
of the polynomial algebra
\[ \ZZ [u_0,...,u_n,q_1,...,q_n,c_1,...,c_{n+1}] \]
by the ideal of relations obtained by equating the coefficients of the
following polynomials in $\gl $:}
\[ \det (A_n + \gl) = \gl ^{n+1} + c_1\gl ^n + ... + c_n\gl + c_{n+1} .\]

\bigskip
In other words, it is the free polynomial algebra in $u$ and $q$ but the
subalgebra of Chern classes, instead of symmetric functions of $u$,
consists of their ``quantum deformations'' from the previous theorem ---
first integrals of the Toda lattice.

\bigskip
Now we can figure out, why one might {\em a priori} expect quantum cohomology
of flag manifolds to be related with at least {\em some} integrable system.

According to our general theory (see $3.8$), equivariant quantum
cohomology of a compact Kahler $U$-manifold $X$ is an algebra of functions on
a lagrangian subvariety $\call $ in a Poisson manifold with $U$-characteristic
classes in the role of Casimir functions. Poisson structure lives in the space
with coordinates
($ q_1,...,q_n$, $p_1,...,p_n$, $c_1,...,c_{n+1}$)
and is given by the formula
\[ q_1 \frac{\p }{\p p_1} \w \frac{\p }{\p q_1} + ... +
   q_n \frac{\p }{\p p_n} \w \frac{\p }{\p q_n} \]
so that the symplectic leaves $\vec{c}=const $ are in fact all isomorphic
to the cotangent bunle of the $q$-torus described above.

Our point now is that although equating Chern classes to non-zero constants
makes little ``cohomological'' sense, the ideal of $\call $
is {\em a priori} a Poisson ideal, and therefore intersections of $\call $
with the symplectic leaves can be interpreted as a $\vec{c}$-parametric family
of lagrangian submanifolds in {\em the same} symplectic space
--- the cotangent bundle of the torus.

Moreover, since the ideal of relations is generated
by quasi-homogeneous $q$-deformations
of the classical relations $c_i=\gs _i(u)$, equations of the lagrangian
submanifolds have the following {\em triangular} form
\[ c_i=C_i(u,q,c_1,...,c_{i-1}),\ i=1,...,n+1 \]
and can be resolved with respect to $c_i$ as
$ c_i=c_i(u,q)$.

This means that the lagrangian submanifolds fit nicely into the phase space
as leaves of a lagrangian foliation --- common levels of the
functions $c_i(u,q), i=1,...,n+1$, which are therefore in involution, ---
and the lagrangian variety $L$ is a singular zero leaf of this foliation.


\bigskip

Our description of quantum (equivariant) cohomology of flag manifolds would be
incomplete without a formula for the intersection pairing (see $3.4$)
\[ \lan \cdot , \cdot \ran :\   QH_{U_{n+1}}^*(F_{n+1},\CC )\otimes _{\CC [c]}
QH_{U_{n+1}}^*(F_{n+1},\CC )\to \CC [c] .\]
Denote $\gS _i(u_0,...,u_{n},q_1,...,q_n)$, $i=1,...,n+1$, the quantum
deformation of elementary symmetric functions $\gs _i(u)$ from Theorem $1$
(\this the
first integrals of the Toda lattice). Let $\gf , \psi \in \CC [u,q,c]$ be two
polynomials considered as representatives of cohomology classes from
$H^*_{U_{n+1}}(F_{n+1})$.

\bigskip
{\bf Theorem 3.}
\[ \lan [\gf ], [\psi ] \ran (c,q)= \frac{1}{(2\pi i)^{n+1}}
\int \frac{\gf (u,q,c) \psi (u,q,c)
du_0 \w ... \w du_n}{(\gS _1(u,q)-c_1)...(\gS _{n+1}(u,q) -c_{n+1})} .\]

\bigskip
The integral here can be replaced by the total sum of $(n+1)!$ residues in the
$u$-space. In order to obtain the intersection pairing in non-equivariant
cohomology $QH^*(F_n)$ it suffices to put $c_1=...=c_{n+1}=0$ in this formula.

Consider the basis $p_1,...,p_n$ of non-negative $(1,1)$-classes on $F_{n+1}$,
$u_i=p_i-p_{i+1}$. Then $(z,p)=z_1p_1+...+z_np_n$ with $z_i > 0$ is
represented by a Kahler form, and $\exp (z,p)$ can be considered as a
non-homogeneous differential form whose degree $(k,k)$ term measures
$k$-dimensional Kahler volume. The corresponding {\em quantum generating
volume function} (see $2.3$):
\[ V(z,q)= \frac{1}{(2\pi i)^n} \int
\frac{ \exp (z,p)\, dp_1\w ... \w dp_n}{\Pi _{j=1}^n (\gS _{j+1}(u(p),q))} \]
has the geometrical meaning of the total Kahler volume of the `$q$-weighted'
space
\[ \calm = \cup _d q^d \calm _d \]
of holomorphic maps $\CC P^1\to F_{n+1}$ of all degrees $d$. The volume is
computed in fact with respect to the Kahler form induced by $(z,p)$ on the
{\em loop space} $LF_{n+1}$ where $\calm $ can be naturally embedded.
Combining our conjectures about general properties of quantum cohomology with
the `conditional' Theorem $3$ we come to the following `unconditional'
prediction.

\bigskip
{\bf Conjecture.} {\em Kahler volume of the space  of parametrized
rational curves of degree $d=(d_1,...,d_n)$ with respect to the Kahler form
with periods $z_1,...,z_n$ on the flag manifold $F_{n+1}$ equals}
\[ \text{Vol}_z (\calm _d) =
\frac{1}{d_1!...d_n!} (\frac{\p }{\p q_1})^{d_1} ...
(\frac{\p }{\p q_n})^{d_n}\, |_{q=0}\, V(z,q) .\]

\bigskip
At $d=0$ this formula reduces to the total volume of the flag manifold itself
and coincides with the fundamental anti-invariant of the permutation group.
The equivariant analogue $V_G(z,q,c)$ of the generating volume function at
$q=0$, $c=\gs (x_0,...,x_n)$ turns into the asymptotic character of
irreducible
representations of $G=U_{n+1}$ with `large highest weights' proportional to $z$
(it can be found using Duistermaat -- Heckmann formula [AB]). It would be
interesting to figure out the meaning of such generating volume functions with
non-zero $q$ and the role of Toda lattices in representation theory
of loop groups. The last question seems to be closely related to
the recent paper [FF] on Toda Field Theory.

\bigskip
{\em Structure of this paper.} In Section $2$ we give a more detailed
review of quantum cohomology theory. Although one can find a number of
approaches to the general theory in the available literature (see for instance
[W] or a recent preprint [S] where in particular the quantum cohomology of
$F_3$ has been computed), we hope that our point of view
is up to certain extent complementary to them. It also should help to clarify
our construction of equivariant quantum cohomology (Section $3$) as well as
those conjectures about its general properties which we exploit in our
inductive proof (Section $4$) of the theorems formulated in this Introduction.

\bigskip
{\em Conventions.} Throughout this paper, we will assume for convenience
that all dimensions are counted in complex units, and --- for the sake
of simplicity --- that all considered compact Kahler
manifolds are simply-connected.

\bigskip
{\em Thanks.}
We would like to express our sincere gratitude to all  participants
of the seminar on mirror symmetry at the Department of Mathematics at UC
Berkeley for their stimulating enthusiasm, and especially to Dmitry Fuchs,
Dusa McDuff, Nikolai Reshetikhin, Albert Schwartz, Vera Serganova and Alan
Weinstein for numerous instructive discussions.

\section{Quantum cohomology \protect\newline and Floer homology}

The objective of this section is to interpret Vafa's construction of quantum
cohomology of a compact Kahler manifold as Floer homology of its loop
space (to be more precise --- of the universal covering of the loop space)
provided with multiplication induced by composition of loops.

\subsection{Additive structure}

\bigskip
Let $X$ be a compact manifold provided with a complex structure $J$ and
a riemannian metric $(\cdot ,\cdot )$ compatible with the complex structure
in the sense that the differential form $\go =(J\cdot ,\cdot )$ is symplectic.

The space $LX$ of contractible (say, smooth) loops $S^1\to X$
inherits from $X$ the same structures:
\begin{itemize}
\item the complex structure $\calj$ which transforms a tangent vector ($=$ a
vector field $t\mapsto v(t)$ along the loop $t\mapsto \gc (t)$ ) to
$t \mapsto J(\gc (t))v(t)$;

\item the $\calj$-compatible riemannian and symplectic forms
  \[ (v,w) = \oint (v(t),w(t))\, dt \ ,\ \gO (v,w)=\oint \go (v(t),w(t))\,dt
;\]

\bigskip
\noindent and additionally carries

\item the action of the reparametrization group $Diff(S^1)$ and in particular
the circle action generated by the vector field $V: \gc \mapsto \dot{\gc }$
on $LX$; and

\item the {\em action functional} $\cala : \tilde{LX} \to \RR $:
\[ \cala (\gc )= \int _D \gf ^* \go \]
which assigns to a loop $\gc $ the symplectic area of a disk
$(\gf: D\to X : \gf |_{S^1=\p D} = \gc )$ contracting the loop, and thus
is well defined only on the universal covering of $LX$.

\end{itemize}

There is a remarkable relation between these structures, namely
\ben
\item the circle action is hamiltonian with respect to the symplectic form
$\gO $ and the hamilton function is $\cala $;
\item the gradient vector field of the action functional relative to the
riemannian metric equals $\calj V$ and thus the gradient ``flow'' consists in
analytic continuation of loops from the unit real circle $S^1\subset \CC - 0$
to its neighborhood in the complex circle.
\een

\bigskip
By definition, Floer homology $FH_*(X)$ is Morse-theoretic homology of the
loop space $LX$ constructed by means of the ``Morse function'' $\cala $ in
the spirit of Witten's approach [W2] to the Morse theory, \this using
{\em bounded} gradient trajectories joining critical points.

historically Floer homology has been introduced [F1] in order to prove Arnold's
symplectic fixed point conjecture and deals with Morse theory of action
functionals perturbed by a hamiltonian term. However the homology itself
is simpler to compute for the unperturbed action functional $\cala $.

In fact the functional $\cala $ is a {\em perfect Morse--Bott--Novikov}
function
on $LX$.

Here
\begin{itemize}
\item ``Novikov'' means that it is multiple-valued and thus the
Morse--Smale complex should be constructed from the critical points on a
covering $\tilde{LX} $ and treated as a module over the group
of covering transformations.
\item The critical points are in fact constant loops
and thus the critical locus of $\cala $ on the covering consists of copies
of the manifold $X$ itself duplicated as many times as many elements are in
the covering transformation group. The critical components are
transversally non-degenerate so that $\cala $ is a Morse--Bott function.
\item The group of covering transformations is in fact the lattice
$\ZZ ^k = \pi _2(X)\cap H_2(X, \RR )$ of spherical periods of closed $2$-forms
on $X$ and thus the Morse--Smale--Bott--Novikov complex can be identified with
the homology group $H_*(X,\ZZ [q,q^{-1}])$ of $X$ where the coefficient ring
is a group ring of the lattice (in the first approximation it can be taken as
the ring of Laurent polynomials in $k$ generators $q=(q_1,...,q_k)$).
\item Finally,
``perfect'' means that the boundary operator in the complex is {\em zero}
so that $FH_*(X)$\newline $\cong H_*(X,\ZZ [q^{\pm 1}])$
as a $\ZZ [q^{\pm 1}]$-module.
\end{itemize}

The latter statement is due to the fact that $\cala $ is the Hamiltonian of a
circle action. The Atiyah convexity theorem [A] says in particular that the
Hamiltonian of a torus action on a compact symplectic manifold is a perfect
Morse--Bott function. A ``scientific'' explanation [G] is that the same
manifold is
the critical set of a function (which leads to the Morse inequality) and the
fixed set of a sircle action (which leads to the opposite Smith inequality
in equivariant cohomology, see also [G1] where locally hamiltonian torus
actions
are considered). A geometrical argument behind this property works pretty well
in the infinite-dimensional Morse theory if one deals with only bounded
trajectories of the gradient flow.

\bigskip

Now we can describe geometrically the Morse--Bott cycles of Floer homology
theory. They are enumerated by ordinary cycles in the components of the
critical
locus. Pick such a component $X$ and a cycle $a\subset X$. The corresponding
Morse--Bott (co)cycle $A\subset LX$ is the union of all the gradient
trajectories
outgoing (resp. ingoing) the critical set $a$ when time $\to -\infty $
($+\infty$ respectively). Since the gradient flow of $\cala $ consists in
analytic continuation, we come to the following description of the cycle $A$:
\begin{quotation}
$ A =\{ $ boundary values of holomorphic maps of the unit disk
   $D\subset \CC $ to $X$ with the center in $a\subset X \ \}$.
\end{quotation}

\subsection{Multiplication}

After such an informal description of the additive structure in Floer homology
it is time to discuss multiplication. There are at least two reasons why
analogue of usual cup-product may not exist in Floer's theory:
\ben
\item intersections in general position of Morse--Bott cycles in $LX$ which
have ``semi-infinite'' dimension would give rise to the cycles of finite
dimension rather than to ``semi-infinite'' cycles again;
\item finite-dimensional Novikov's cohomology is cohomology with local
coefficients determined by periods $\log q$ of the closed $1$-form; cup-product
of such cohomology is accompanied by tensor multiplication of the local
coefficient systems and would give rise to $q^2$ in the product, instead of
$q$ again.
\een

\bigskip
In fact the multiplicative structure in Floer homology is analogous to the
convolution in the homology of a Lie group induced by multiplication in the
group. The ``group'' operation on $LX$ consists in composing parametrized
loops at the marked point $t=0$ on the circle $S^1$. This operation is
ill-defined since the loops we consider are {\em free}. However this operation
considered as a {\em correspondence} can be described by its graph in $LX^3$,
and the convolution multiplication $A*B$ of Morse--Bott cycles can be defined
through intersection indices $\lan A*B,C\ran $ of the products
$A\times B\times C \subset LX^3$ with the graph.

By some technical analytical reasons it is more convenient to perturb the graph
and consider instead the cycle in $LX^3$ which consists of triples of loops
which are boundary values of a {\em holomorphic map of ``pants''} to $X$.
More generally, one can define multiple products $A_1*...*A_N$ through
intersection indices $\lan A_1*...*A_N, C\ran $ in $LX^{N+1}$ considering
compositions of pants and their holomorphic maps to $X$.

In more detail, denote $\Pi _N$ the standard Riemann sphere $\CC P^1$ with $N$
disks detached and their boundaries {\em left} oriented and parametrized by the
standard unit circle $S^1$. Denote $\gC _N$ the cycle in $LX^N$ which consists
of $N$-tuples of boundary values of holomorphic maps $\Pi _N \to X$. For $N$
given Morse--Bott cycles $A_1,...,A_N$ in $LX$ define their
$\lan A_1|...|A_N \ran$ as the intersection index
of ``semi-infinite cycles'' $A_1\times ... \times A_N \subset LX^N$ and $\gC
_N$.

We should make a correction here: the intersection index should be defined as
{\em Novikov's} one. This means that the product
$A_1\times ... \times A_N$ should be considered as a cycle on the diagonal
$\ZZ ^k$-covering $(LX^N)^{~}$. An important property of $\gC _N$ is that
it has a canonical lifting to this covering: an $N$-tuple of the boundary
values is provided with the homotopy type of the map $\Pi _N \to X$.
Novikov's intersection index of two {\em transversal} cycles $A$ and $B$
on the covering, by definition,
assumes values in the group ring of the covering and counts {\em isolated}
intersection points of the cycles projected to the base, with signs and
``weights'' $q^d\in \ZZ[q^{\pm 1}]$, where $d\in \ZZ ^k$ is the covering
transformation that transforms the preimages in $A$ and $B$ of the intersection
point into one another.

Now we can describe geometrically an intersection event of
$A_1 \times ... \times A_N$ with $\gC _N$. The Morse--Novikov cycles
$A_i$ correspond to some finite-dimensional cycles $a_i$ in $X$. An
intersection point, on one hand, is an $N$-tuple of loops which are boundary
values of $N$ parametrized holomorphic disks in $X$ with centers respectively
in $a_1, ..., a_N$. On the other hand it is the $N$-tuple of boundary values
of a holomorphic map $\Pi _N \to X$. Due to the uniqueness of analytic
continuation, the disks and $\Pi _N$ glue up to a single holomorphic map
$\gf : \CC P^1 \to X$ with the centers $x_1,...,x_N$ of the (formerly detached)
disks
being mapped to the cycles $a_1,...,a_N$ respectively. The group element $d$
in the definition of Novikov's intersection index, in our situation
measures the difference of homotopy types of the two holomorphic films attached
to the $N$-tuple of loops and equals the homotopy type of the map $\gf $, \this
the degree of the rational curve $\gf (\CC P^1)$. Thus we come to Vafa's
formula:
\[ \lan A_1|...|A_N\ran =
\sum _{\begin{array}{c} \text{isolated holomorphic maps} \\
                    \gf : (\CC P^1,x_1,...,x_N)\to (X,a_1,...,a_N) \end{array}}
\pm q^{\deg \gf }   .\]

The assumption that the intersected cycles are transversal means that the
number of independent holomorphic sections of the induced tangent bundle
$\gf^! T_X$ equals the Euler characteristic $c(d)+\dim X$ prescribed by the
Riemann--Roch formula, and the constraints $\gf (x_i)\in a_i$ are
non-degenerate
(in the sense of implicit function theorem). Thus the isolatedness
implies
\[ c(d)+\dim _{\CC } X = \sum _i \codim _{\CC } a_i .\]
Notice that holomorphic spheres constrained at two points are never isolated
(circle action! By the way it is that geometrical argument that makes $\cala $
perfect) and thus the double intersection index $\lan A, B\ran $
coincides with the non-degenerate Poincare pairing of cycles $a,b$ in $X$.
One can identify a cycle $a$ of codimension $\ga $ in $X$ with the
Poincare-dual cohomology class of degree $\ga $. The above formula means that
$\lan A_1|...|A_N\ran $ defines in this way a ``quantum'' $q$-valued
intersection pairing $H^*(X)^{\otimes N} \to \ZZ[q^{\pm 1}]$ which respects
the usual grading in cohomology provided that $\deg q^d = c(d)$:

\[ \deg \lan a_1|...|a_N\ran = \deg _{\CC} a_1 +...+\deg _{\CC } a_N
- \dim _{\CC} X .\]

The triple ``pairing'' can be used in order to define the ``quantum
multiplication'' $a*b$:
\[ \forall c \ \ \lan a*b, c\ran = \lan a|b|c \ran .\]
The fact that this multiplication is associative as well as that the multiple
pairings can be expressed through $*$-operation and Poincare pairing with the
fundamental cycle $\bf [1]$ as
\[ \lan a_1|...|a_N\ran = \lan a_1*...*a_N, \bf 1 \ran ,\]
reduces to the principal axiom of Topological Field Theory:
i\begin{quote}
If the surface $\Pi _N$ is cut by a circle into a union of two surfaces
$\Pi _{M+1}$ and $\Pi _{N-M+1}$ then the corresponding intersections satisfy
\[ \lan a_1|...|a_N\ran =\sum _j \lan a_1|...|a_M|b_j\ran
                                 \lan c_j|a_{M+1}|...|a_N\ran \]
where $\sum _j b_j\otimes c_j \in H^*(X\times X)$ is Poincare-dual to the
class of the diagonal $X\subset X\times X)$.
\end{quote}

\bigskip
Rigorous justification of this axiom as well as of correctness of the above
definitions is obstructed by a number of highly non-trivial problems.

First of all, in order to bring the cycles in $LX^N$ to transversal position
one needs, in general, to perturb the complex structure on $X$ toward almost
complex structures, and the whole story begins to depend on Gromov's theory
[Gr] of pseudo-holomorphic curves in symplectic manifolds and compactifications
of their moduli space.

Even in the additive Floer theory some difficulties (with multiple
coverings of holomorphic curves) has not been overcome so far. The situation
seems to be simpler, and the difficulty --- resolved, in the case of almost
Kahler manifolds with positive first Chern class $c$ and almost complex
structure close to an integrable one (see [O]). The case of zero first
Chern class which also has been worked out [HS], requires Novikov's
completion of the group ring $\ZZ [q^{\pm 1}]$ (Vafa's formula may contain
infinite sums).

In the cases when the additive theory can be completed successfully,
correctness of the definitions of multiple intersection indices, their
skew-commutativity, independence on moduli of surfaces $\Pi _N$, on the choice
of cycles in the homology classes, and so on, does not seem to exhibit
further complications (see [R]).

At the same time, associativity of the quantum multiplication and the axioms
of Topological Field Theory have been verified, as far as we know, only in the
simplest case of manifolds $X$ with $\pi _2(X)=0$ (M.Schwartz) where instanton
corrections do not occur at all.

\subsection{Alternative approaches}

We briefly review here some other constructions of quantum cohomology
algebras. Later they will be described in more detail in connection with
equivariant theory.

\bigskip
First of all, instead of the ill-defined composition map $LX\times LX \to LX$
one can consider a well-defined evaluation map $LX \to LX\times X$:
\[ (\text{a loop}\ t\mapsto \gc (t)) \mapsto (\gc \in LX, \gc (t_0)\in X) .\]
It induces a linear map
\[ H^*(X)\otimes FH^*(X) \to FH^*(X) \]
and thus makes cohomology classes of $X$ act on the Floer cohomology
$H^*(X, \ZZ [[q^{\pm 1}]])$ of the loop space by $\ZZ [[q^{\pm 1}]]$-linear
operators. These operators, along with operators of multiplication by $q$,
generate some associative skew-commutative operator algebra. Composition
of such operators differs in fact from ordinary cup-product in $H^*(X)$.
It is  not obvious from this point of view even that they should form an
algebra
closed with respect to composition. However interpretation of matrix
elements of such operators in terms of rational curves in $X$ leads directly
to Vafa's definition of quantum cup-product. Such a module structure in
Floer homology of $LX$ over
cohomology of $X$ itself has been exploited many times in the literature
on symplectic topology [FW], [Oh], [F2], [H], [G1], [G2] (and in a recent
paper [S] on quantum cohomology).

\bigskip
A similar approach, based however on differential forms, was studied in [V].
A closed differential $r$-form $p$ on $X$ and a density $\gr $ on the unit
circle determine a closed differential $r$-form $P$ on the loop space $LX$:
\[  P|_{\gc }(v_1, ... , v_r)  =
     \oint p|_{\gc (t)} (v_1(t),...,v_r(t)) \gr (t) dt .\]
The ordinary cohomology class of $P$ on $LX$ depends, by the Stokes theorem,
only on the class of $p$ on $X$
and on the total ``mass'' $\oint \gr (t) dt$.
However we are going to integrate $P$ over non-compact cycles in $LX$,
so that the Stokes theorem does not apply literally. The cycle we need is
denoted $\calm _d$ and consists of {\em algebraic loops of degree} $d$ in $X$,
\this degree $d$ holomorphic maps $\CC P^1 \to X$ which can be considered as
elements of the loop space if we restrict them to the unit circle in $\CC -0$.
The cycle $\calm _d$ --- a ``moduli space'' of rational curves --- can be
compactified, after Gromov [G], by reducible curves, and this is a reason
to expect that the integral converge. The reducible curves however do not
correspond to any loops, and the compactification can not be done inside $LX$.

One can define quantum intersection pairings as
\[ \lan p_1|...|p_N \ran = \sum _d \pm q^d
              \oint _{\calm _d} P_1\w ... \w P_N .\]
assuming the corresponding densities $\gr _i,\ i=1,...,N$, being of unit total
mass each and generic. The integrals in this sum can be non-zero only if
the total degree $r_1+...+r_N$ of the differential form equals the dimension
$2(c(d)+\dim X)$ of the cycle $\calm _d$ and reduces to
$\int _X p_1\w ... \w p_N$ for $d=0$.

\bigskip
The coincidence of such intersection pairings with previously defined ones
becomes ``obvious'' if we interpret them in the spirit of integral geometry.
Imagine that the densities $\gr _i$ has been chosen as Dirak $\gd $-functions
concentrated at $N$ generic marked points $x_1,...,x_N$ on the unit circle.
Then
\[ \int _{\calm _d} P_1\w ... \w P_N =
\int _{\bar{\calm _d}} \bar{p}_1\oplus ... \oplus \bar{p}_N  \]
where $\bar{p}_i$ is a differential form on $X^N$ obtained as the pull-back of
of $p_i$ on the $i$-th factor, and $\bar{\calm _d}$ is the closure in $X^N$ of
the image of the evaluation map
\[ (\calm _d \subset LX) \to X^N:\ \gc \mapsto (\gc (x_1), ... ,\gc (x_N) ) .\]
The fundamental class of the complex variety $\bar{\calm _d}$ in $H^*(X^d)$
is the same for {\em generic} marked points. Taking the average value of such
integrals, defined by means of $\gd $-densities, over the torus $(S^1)^N$ in
the configuration space $(\CC -0)^N$ of marked points we conclude that the
quantum intersection pairing of closed forms depends only on their cohomology
classes in $H^*(X)$ and does not depend on the densities provided that they
are,
say, continuous. On the other hand, replacing the forms $p_i$ by their
Poincare-dual cycles we find the integral equal to an intersection index
in $X^N$  with the ``moduli space'' $\bar{\calm _d}$, and this leads back
to the original Vafa's construction --- counting rational curves constrained
at marked points. Notice that this construction of $\lan p_1|...|p_N \ran$
as intersection indices in $X^N$ also explains how the signs in Vafa's
formula should be chosen.

\bigskip
The last construction of quantum cohomology algebras --- via generating volume
functions --- is most convenient in the case when the ordinary cohomology
algebra $H^*(X)$ is generated (as an algebra) by Kahler classes, and will be
described below
under this assumption. Let $p_1,...,p_k$ be an integer basis of non-negative
$(1,1)$-forms in $H^2(X)$, $p(z)=z_1p_1+...+z_kp_k$ be a general linear
combination. If $p(z)$ is a Kahler form on $X$ the corresponding form $P(z)$
is a Kahler form on the loop space $LX$, and the following formal series
 \[ V(z,q)=\sum _d q^d \int _{\calm _d} \exp (P(z)) \]
represents the Kahler volume of the ``weighted moduli space''
\[ \calm = \cup _d \ q^d\calm _d ,\]
since the terms of the exponential series
\[ \exp P = \sum _r \frac{1}{r!} P\w ... \w P \ \ (r\, \text{times}) \]
represent $r$-dimensional Kahler volumes with respect to $P$.

We call $V(z,q)$ {\em generating volume function} (in fact it is a simplified
version of the generating correlation function $\Phi $ from CTFT [W],[D],[K]).

It has the following properties:
\ben
\item $V(z,q)$ becomes quasi-homogeneous of degree $-\dim X$ if we put
$\deg z_i=-1$, $\deg q_i=D_i$ where $c=D_1p_1+...+D_kp_k$ represents the
$1$-st Chern class of $X$ in the basis $(p_1,...,p_k)$;
\item $ V(z,0) = \int _X \exp (p(z)) $ is the volume function of $X$;
\item quantum intersection indices
of the generators $p_1,...,p_k$ can be expressed in terms of $V(z,q)$ as
\[ \lan p_{i_1}|...|p_{i_N}\ran =
     \frac{\p ^N}{\p z_{i_1} ... \p z_{i_N}} |_{z=0} V(z,q) \]
(this is due to the very property of the exponential function).
\een

This last formula implies that one can {\em define} the quantum cohomology
algebra $QH^*(X)$ as the quotient of the polynomial algebra $ \ZZ [p,q] $
by the ideal $I$ of all polynomials $R(p,q)$ such that
\[ R(\p /\p z_1,...,\p /\p z_k, q_1,...,q_k) V(z,q) = 0 .\]

\bigskip
{\bf Example: $QH^*(\CC P^1)$.} A holomorphic map $\CC P^1 \to \CC P^1$
of degree $d$ is given by the ratio $f/g$ of two homogeneous polynomials
\[ f=\sum a_i x^iy^{d-i}, \ g=\sum b_i x^iy^{d-i} \]
 in two variables. This means that the space $\calm _d$ of such maps
compactifies to the complex projective space $\CC P^{2d+1}$. Let $p$ be the
Fubini Kahler form on the target $\CC P^1$. It is obtained from the form
\[ \p \bar{\p } \log (f\bar{f} + g\bar{g} ) \]
in homogeneous coordinates $(f,g)$. The corresponding Kahler form $P$ on
$\calm _d \subset L(\CC P^1)$ is similarly obtained from
\[  \p \bar{\p } \log (f\bar{f} + g\bar{g} )|_{(x,y)=(e^{it},1)} \]
as their mean value over $t$. At $t=0$ this gives
\[ \p \bar{\p } \log [ |\sum a_i|^2 + |\sum b_i|^2 ] \]
and leads to a non-negative $(1,1)$-form which extends to $\CC P^{2d+1}$ and
represents there a generator of $H^2(\CC P^{2d+1})\cong \ZZ $. The same
properties hold for all $t$, and thus $P$ represents the class the of Fubini
form on $\CC P^{2d+1}$. We conclude that
\[ V(z,q) = \sum _{d=0}^{\infty } \frac{z^{2d+1}}{(2d+1)!} q^d .\]
It is easy to see that the ideal $I$ of polynomials $F(\p /\p z, q)$
annihilating $V$ is generated by $(\p /\p z)^2 - q$ and therefore
\[ QH^*(\CC P^1) = \ZZ [p,q]/(p^2-q) .\]
We find {\em a posteriori} that it is indeed a $q$-deformation of the
classical cohomology ring $H^*(\CC P^1)=\ZZ [p]/(p^2) $.

\subsection{Characteristic lagrangian variety}

Keeping the assumption, that cohomology algebra of $X$ is generated by
Kahler classes, and the notations introduced in the end of $2.3$, we
describe here $QH^*(X,\CC )$ as the algebra of functions on some lagrangian
variety.

Since the quantum cohomology algebra is now identified with the quotient
$\CC [p,q]/I$, its spectrum is a subvariety $L$ in the space $\CC ^{2k}$ with
coordinates $(p_1,...,p_k,q_1,...,q_k)$ with the ideal $I(L)=I$ (strictly
speaking, the variety can be defined only over formal series if the $1$-st
Chern class $c$ of $X$ is not positive). In any case, it is quasi-homogeneous
with $\deg p_i=1, \deg q_i=D_i$. The space $\CC ^{2k}$ has the canonical
Poisson structure
\[ \sum _{i=1}^k q_i \frac{\p }{\p p_i} \w \frac{\p }{\p q_i} \]
which is nothing but extension of the canonical symplectic structure
\[ \sum dp_i\w \frac{dq_i}{q_i} \]
on the cotangent bundle
\[ T^*B=H_2(X,\CC)\times [H^2(X,\CC)/2\pi \sqrt{-1}\ZZ ^k] \]
of the torus $B$ dual to the $2$-nd homology lattice $\ZZ ^k$.
We claim that the variety $L$ is lagrangian with respect to this symplectic
form.

Indeed, interpret the Floer cohomology space
\[ FH^*(X,\CC )=H^*(X,\CC)\otimes \CC [q^{\pm 1}] \]
as the space of vector-functions of $q$ with values in the vector space
$W=H^*(X,\CC )$ and introduce the following {\em operator-valued} $1$-form
\[ A = \sum A_i(q)\frac{dq_i}{q_i} =
(p_1*)\frac{dq_1}{q_1} + ... + (p_k*)\frac{dq_k}{q_k} .\]
Here $A_i=p_i*$ is understood as the operator on $W$ of quantum multiplication
by $p_i$ computed at a particular value of $q$.
First of all, we claim that this $1$-form satisfies:
\[ A\w A=0 , \ dA=0 \]
(which means in fact that $\gel d + A\w $ is a flat connection operator for all
$\gel $).
The $1$-st identity simply means that the operators $A_i$ commute so as $p_i*$
do.
The $2$-nd identity means that the matrix elements of $A$ are closed $1$-forms
and does not follow from any formal properties of quantum multiplication which
have been discussed so far. It can be reformulated, in terms of matrix
elements of $A_i$, as follows:
\begin{quotation}
For any two cycles $a$ and $b$ in $X$ the quantum intersection indices
$\lan a|p_i|b \ran $ are partial derivatives $q_i\frac{\p S}{\p q_i}$ of a
single (locally defined) function $S=S_{a,b}(q)$.
\end{quotation}

Put
\[ S_{a,b}= \sum _i \lan a,p_i,b \ran \log (q_i) \ + \]
\[ +\ \sum _{\begin{array}{c} \text{ rational curves in}\, X \\
            \text{  with}\, 0\in a, \infty \in b\, \text{ of degree}\, d>0 \\
           \text{and with}\  c(d)+\dim X=\codim _{\CC }a+\codim _{\CC }b +1
\end{array} } \pm q^d .\]
The $1$-st sum is a potential for the constant terms in $\lan a|p_i|b\ran $
and involves classical intersection indices. The $2$-nd sum counts non-constant
rational curves, constrained at two points, as if they were discrete. If such a
curve contributes by $\pm q^d$ to $S_{a,b}$ then it contributes by
$\pm d_iq^d$ to $q_i\p S_{a,b}/\p q_i$. Here $d_i$ is exactly the intersection
index of a complex hypersurface Poincare-dual to $p_i$ with this rational
curve. This means that there are exactly $d_i$ ways to parametrize the curve in
such a fasion that $0\in a$, $\infty \in b$ and $1\in p_i$, and hence the curve
contributes to $\lan a|p_i|b\ran $ with the same weight $\pm d_iq^d$. This
proves our assertion (modulo our usual reservations). In fact this $S_{a,b}$ is
one of the ``higher order'' pairings considered in Conformal
Field Theory (actually it is the lower order pairing).

Now the lagrangian property of $L$ follows from a general lemma
(which we learned from N.Reshetikhin).

\bigskip
{\bf Lemma.} {\em Let
\[   A = \sum _i A_i(t)dt_i \]
be a matrix-valued differential $1$-form satisfying $A\w A=0$ and $dA = 0$.
Let the scalar differential $1$-form
\[ p = \sum _i p_i(t) dt_i \]
be its {\em simple} eigen-value. Then $p$ is closed.}

\bigskip
{\em Proof.} The assumption actually means that the commuting matrices $A_i(t)$
have a common eigen-vectors $w(t)$ such that $A_i(t)w(t)=p_i(t)w(t)$. Being
simple, the eigen-vectors can be chosen smooth in $t$, and the transposed
matrices $A_i^*$ have a smooth field of eigen-covectors $w^*(t)$ (with the same
eigen-values) normalized in such a way that $\lan w, w^* \ran =1$ identically.
Now we have
\[ d(pdt) = d (\lan w,w^*\ran (pdt)) = d \lan Aw,w^* \ran = \]
\[ \lan (dA)w,w^*\ran - \lan A\w dw,w^* \ran   - \lan Aw, dw^* \ran = \]
\[ \lan dw, A^*w^* \ran -\lan Aw, dw^* \ran =(d\lan w,w^*\ran )\w (pdt)=0 .\]

\bigskip
Applied to our quantum cohomology situation, this lemma shows that every
non-singular local branch of $L$ over $B$ is a lagrangian section of $T^*B$.
This implies that $I$ is a Poisson ideal at least in the case if $I=\sqrt{I}$.

\bigskip
Below we explain how intersection pairings and generating volume functions
can be described in terms of geometry on $L$ assuming for simplicity that
$I=\sqrt{I}$ and that the $1$-st Chern class of $X$ is positive (so that $L$
is indeed a quasi-homogeneous affine algebraic subvariety in $\CC ^{2k}$ with
coordinates $(p,q)$).

Consider the class in quantum cohomology algebra of $X\times X$ Poincare-dual
to the diagonal $X\subset X\times X$. It can be considered as a function on
the characteristic lagrangian variety of $X\times X$ which is nothing but
$L\times L$. Restrict this function to the diagonal $L\subset L\times L$
and denote the restriction $\gD\in \CC[L]$. Let $\gf _1,...,\gf _N \in \CC [L]$
be some quantum cohomology classes. Then for generic $q\in B$
\[ \lan \gf _1|...|\gf _N \ran (q) =
      \sum _{p\in  L\cap T^*_qB} \frac{\gf _1(p)...\gf _N(p)}{\gD (p)} \]
and
\[ V(z,q)=\sum _{p\in L\cap T^*_qB} \frac{\exp (z_1p_1+...z_kp_k)}{\gD (p)} .\]

The last remark: since $L$ is lagrangian, the action $1$-form on $T^*B$
restricted to $L$ is exact,
\[ \sum p_i\frac{dq_i}{q_i} |_{L} = dC, \ \ C\in \CC [L] .\]
Using quasi-homogeneity of $L$ and Cartan's homotopy formula one can easily
show that $C=D_1p_1+...+D_kp_k$ is the $1$-st Chern class of $X$ understood
as a function on $L$.

\section{Equivariant quantum cohomology}

\subsection{Why ``equivariant''?}

In our inductive computation of quantum cohomology of flag manifolds we will
encounter the following kind of problems. With a vector bundle over some
base $B$ one can associate a fiber bundle $E\to B$ whose fibers are flag
manifolds --- they consist of flags in the fibers of the vector bundle.
Consider the maps of $\CC P^1$ with $N$ marked points to $E$ whose composition
with the projection to $B$ maps $\CC P^1$ to a point
and which are holomorphic if considered as maps to the fiber flag
manifolds. We will call such holomorphic curves {\em vertical}.

One may pick $N$ cycles in $E$ and ask how many of such vertical
parametrized rational curves of certain homotopy type have the $1$-st marked
point on the
$1$-st cycle, the $2$-nd marked point --- on the $2$-nd cycle, and so on.

When the base $B$ is a point, the problem (properly understood of course in
terms
of intersection indices) becomes a question about structural constants of the
quantum cohomology algebra of the flag manifold. Our more general problem about
rational curves in flag bundles will not arise in its full generality
--- we will rather need a sequence of special bundles of flag manifolds over
Grassmannians and holomorphic hypersurfaces in the role of the cycles.

On the other hand, this sequence of problems can be understood better in the
context of vector bundles over arbitrary finite cellular bases since in such
generality it can be
replaced by a {\em universal} problem about the universal vector bundle over
the classifying space $BG$. The total space of the universal flag bundle
$E\to BG$ is nothing but the homotopic quotient $EG\times _G F$ of the
flag manifold $F$ by the unitary group $G$. Therefore our universal
problem reduces to the question about structural constants of what should be
called the {\em equivariant} quantum cohomology algebra of the flag manifold.

\subsection{``Classical'' equivariant cohomology}

Recall some standard facts [Hs], [AB] about equivariant cohomology.

Let $X$ be a manifold provided with a left action of a compact Lie group $G$.
Consider the {\em universal} principal $G$-bundle $EG\to BG$ --- a principal
$G$-bundle with contractible total space $EG$, and define the {\em homotopic
quotient} $X_G$ of $X$ by $G$ as $EG\times _G X = (EG\times X)/G$.

\bigskip
{\bf Examples.} 1) If $X$ is a point then $X_G=EG/G=BG$.

2) If $H\subset G$ is a Lie subgroup, $X$ is the homogeneous space $G/H$ then
$(G/H)_G=EG\times _G(G/H)=(EG\times _G G)/H=EG/H=BH$. For instance, if $G$ is
the unitary group $U_n$ and $H$ is its maximal torus $T^n$ so that
$X$ is the flag manifold $F_n$ then $X_G=BT^n=(\CC P^{\infty})^n$.

\bigskip
The equivariant cohomology $H^*_G(X)$ of a $G$-space $X$ is defined as the
ordinary cohomology $H^*(X_G)$ of its homotopic quotient. The natural
fibration $X_G \to BG$ (with fiber $X$), induced by the projection of
$EG\times X$ on the first factor, along with Example 1), provide the
equivariant
cohomology with a module structure over the coefficient algebra $H^*_G(pt)$
of the equivariant theory which is nothing but the characteristic class algebra
$H^*(BG)$ of the group $G$.

\bigskip
{\bf Example.} For the flag manifold $F_n$ its $U_n$-equivariant cohomology can
be identified with the polynomial ring in $n$ generators $(u_1,...,u_n)$
since $H^*(\CC P^{\infty})=\CC [u]$ where $u$ is the $1$-st Chern class of the
universal Hopf circle bundle.  The module structure over the algebra of
universal Chern classes $H^*(BU_n)=\CC [c_1,...,c_n]$ becomes more ``visible''
if we represent the equivariant cohomology of the flag manifold as the quotient
of the polynomial algebra $\CC [u, c]$ by the ideal of
relations $c_i=\gs _i(u)$, $i=1,...,n$, where $\gs _i$ are elementary symmetric
polynomials of $(u_1, ..., u_n)$.

Similarly, equivariant cohomology of cartesian products of flag manifolds
are tensor products of equivariant cohomology of factors and they are
modules over characteristic class algebras of products of unitary groups.
Of course, this is a general property of products $\Pi X_i$ of $G_i$-spaces.

\subsection{Equivariant intersection indices}

Consider a $D$-dimensional compact oriented $G$-manifold $X$ and the
associate $X$-bundle
$\pi : X_G \to BG$. Since we are actually going to apply our general
constructions
to homogeneous complex manifolds it is convenient to make a convention right
now that all the dimensions are {\em complex} ones, and therefore dimensions
of real manifolds or cycles can be {\em half}-integral. With this convention
in force, let us consider  equivariant cohomology classes $p_1,...,p_N$ of
$X$ of total degree $M$ and define their {\em intersection index}
$\lan p_1,...,p_N \ran $ with values
in the structural ring  $H^*_G(pt)$ of equivariant theory.

If $C$ is a homology class of $BG$ of degree $K$ one can construct its {\em
inverse image} $\pi ^{-1}(C)$ which is geometrically the preimage
of the cycle $C$ in the bundle $\pi : X_G \to BG$ and represents a homology
class of degree $K+D$ in $X_G$. By definition,
\[ \lan p_1,...,p_N \ran [C] = (p_1 ... p_N) [\pi ^{-1}(C)] .\]
This formula describes the intersection cohomology class through its
evaluation on homology classes and may give rise to a non-zero result only if
$M=K+D$ of course. In the case when an infinite-dimensional manifold has been
chosen on the role of the classifying space $BG$ one may also think of
$p_1,...,p_n$ as cycles of finite total {\em co}dimension $M$, and of
$\lan ... \ran [C]$ as the mutual intersection index of $p_1,...,p_N$ and
$\pi ^{-1}(C)$. In the case if $C$ is a point our definition reduces
to the ordinary intersection index in $X$ of cycles Poincare-dual to the
restrictions of the cohomology classes $p_i$ to the fiber of $\pi $.

The equivariant intersection indices $H^*_G(X)^{\otimes N} \to H^*_G(pt)$
have the following more or less obvious properties:

\ben
\item They are homogeneous of degree $-\dim X$ (with our
convention in force);
\item They are $H^*(pt)$-multi-linear;
\item They are totally anti-symmetric (notice that $H^*(pt)$ happened to be
commutative);
\item They are determined by cup-multiplication in $H^*_G(X)$ and by the
``intersection index'' $H^*_G(X)\to H^*_G(pt)$ with $N=1$ which is
nothing but the {\em direct image} operation $\pi _! : H^*(X_G)\to H^*(BG)$
dual to the inverse image in homology.

\een
In terms of differential forms the
direct image operation consists in fiberwise integration.

\bigskip
Our objective for the moment is to describe explicitly the direct image for
equivariant cohomology of flag manifolds.

{\bf Proposition.}
{\em For the flag manifold $F_n$ the direct image
$\pi _!: \CC [u] \to \CC [c]$ is given by
the following Cauchy formula:}

\[ (\pi _! f)(c)=  (\frac{1}{2\pi i})^n \int _{T^n} 
\frac{f(u) du_1\wedge ...\wedge du_n}{(\gs _1(u)-c_1)...(\gs _n(u)-c_n)} .\]

\bigskip
The integral equals the total sum of residues in $\CC ^n$. In other words,
in order to find the direct image of a polynomial $f(u)$
one first constructs its total alternation
\[ \text{Alt}f(u)=\sum _{w\in S_n} (-1)^{\gel(w)} f(wu) ,\]
then divides it by the ``fundamental anti-invariant'' ($=$ Vandermond)
\[ \gD _n (u) = \det (\frac{\p \gs _i(u)}{\p u_j}) \]
and expresses the ratio $\text{Alt}f/\gD _n $ as a polynomial
$\hat{f} (\gs (u))$
of elementary symmetric functions: $\hat{f} (c_1,...,c_n)$ is then the direct
image of $f$.

The main argument in the proof of this formula is ``what else can it be?''

Indeed, due to linearity property the direct
image operation is completely determined by its action on generators
of $\CC [u]$ as a $\CC [c]$-module. The generators can be chosen
as homogeneous representatives of a linear basis in the ordinary
cohomology $\CC [u]/(\gs _1(u),...,\gs _n(u))$ of the flag manifold
(Nakayama lemma!). Due to the degree reasons these representatives all
have zero direct images except the generator Poincare dual to the
fundamental cycle. The latter has constant direct image, and the constant
can be easily found equal $1$ (evaluate the direct image at a point).
The residue formula (and the operation $\gD _n ^{-1}\text{Alt}$) do have
all there properties since $\deg \gD _n$ ``accidentally'' equals $\dim F_n$.

\bigskip
One more example. Consider the subgroup
$G'=U_m\times U_{n-m} \subset U_n=G$ and the bundle
$BG'\to BG$ with the fiber $G/G'=Gr(n,m)$. The direct image operation
\[ \text{Direct image}:\ H^*(BG')=\ZZ [c'_1,...,c'_m,c''_1,...,c''_{n-m}] \to
\ZZ [c_1,...,c_n] = H^*(BG) \]
in this bundle somehow transforms  partially symmetric polynomials of \newline
$(u',u'')$ $=$ $((u_1,...,u_m),(u_{m+1},...,u_n))$
to totally symmetric ones, since
\[ c'_i=\gs _i(u'),\ c''_j=\gs _j(u''), \ c_r=
\sum _{i=0}^r \gs _i(u') \gs _{r-i}(u'')  = \gs _r(u) \]
(where $\gs _0=1$).

\bigskip
{\bf Corollary.}
\[ [\text{\em Direct image}\ f](\gs (u)) = \frac{
Alt [ \gD _m(u')\gD _{n-m}(u'')f(\gs (u'),\gs (u''))]}{m! (n-m)! \gD _n(u)} .\]

\bigskip
{\em Proof.} We can represent $f(c',c'')$ as the direct image
 $\Pi _! g(u',u'')$ of some $g(u)$  in the product of bundles
$\Pi : BT^m\times BT^{n-m} \to BU_m\times BU_{n-m}$ and thus identify
$[Direct image f]$ with $\pi _! g$.

\subsection{Instanton corrections}

Let $X$ be a complex Kahler manifold of dimension $D$ provided with a
holomorphic action of the complexified compact Lie group $G\subset G_{\CC}$.
We will assume for simplicity that $X$ that
$H^{1,1}(X)=H^2(X)$.
%
%
Notice that the lattice $\ZZ ^k$ is a sublattice in the second
homology group of the homotopic quotient $X_G$ and thus classes of
{\em vertical}
rational curves in the total space of the bundle $X_G\to BG$ are canonically
identified with elements of $\ZZ ^k$.
%
%

We define {\em  quantum equivariant
intersection indices} as follows.

Let $p_1,...,p_N$ be cycles in $X_G$ of finite codimensions which add up to
$M$.
Their quantum intersection index $\lan p_1 |...| p_N \ran$ will be an element
of the algebra $H^*_G(pt, \ZZ [[q]])$. Given a $K$-dimensional cycle
$C\subset BG$, we define the value $\lan p_1|...|p_N \ran [C]$ as the sum of
contributions of rational parametrized curves $\gf : \CC P^1 \to \pi ^{-1}(C)$
in the {\em fibers} of the bundle $\pi : X_G\to BG$ restricted to $C$ such
that $N$ marked points $x_1,...,x_N$ in $\CC P^1$ map to the cycles
$p_1, ..., p_N$ respectively: $\gf (x_i)\in p_i$. The contribution of $\gf $
is non-zero only if $c(d)+D+K=M$ and equals $\pm q^d$ in the assumptions of
course that the cycles $p_i$ are in
general position with respect to the family of vertical rational curves
$\gf $ of degree $d$, that the family indeed has the dimension $c(d)+D+K$
predicted by the Riemann--Roch formula, and that the contributing curves are
regular points in this family:

\[ \lan p_1|...|p_N \ran [C] =
\sum _{\begin{array}{c} \text{vertical discrete holomorphic maps:} \\
         (\CC P^1,x_1,...,x_N) \to (\pi ^{-1}(C),p_1,...,p_N) \\
                        \text{ of degree} \ d \end{array}}
 \pm q^d .\]

The sign $\pm $ in this formula can be defined naturally in terms of
intersection indices in moduli space; it is ``plus'' at least in the case
if all the cycles $p_i$ and $C$ are holomorphic (the latter assumes that
a complex manifold is taken on the role of $BG$), and will be described
in $3.5$ for arbitrary $C$.

Rigorous justification of this construction, and in particular --- verification
that the intersection indices actually depend only on the (co)homology classes
represented by the cycles $p_i$ and $C$, encounters the same difficulties
as in the case of the quantum non-equivariant intersection indices. In
particular, bringing to general position may involve perturbations of the
complex structure towards almost complex ones which in our case should be
done fiberwise in the bundle $X_G\to BG$ and do not have to be the same on all
fibers.

\bigskip
Intersection indices $\lan \ |...|\ \ran $ have the following obvious
properties
relating them with ``classical'' intersection indices $\lan \ , ... ,\ \ran$:
\ben
\item they are multi-linear and skew-symmetric;
\item $\lan p_1 |...| p_N| [1] \ran = \lan p_1| ... |p_N \ran $, where
$[1]$ represents the fundamental cycle in $X_G$;
\item $\lan p_1|...|p_N\ran |_{q=0} = \lan p_1,...,p_n \ran$ --- they are
$q$-deformations of classical intersection indices;
\item $\lan p_1 | p_2 \ran = \lan p_1, p_2 \ran $ so that $\lan p | [1] \ran $
coincides with the classical direct image operation;

\noindent and a less obvious

\item $H^*_G(pt)$-multi-linearity property (where `$\cdot $' stands for
the cap-product, Poincare dual to the ordinary multiplication of cohomology
classes represented by finite codimension cycles)
\[ \lan \pi ^*(p)\cdot p_1|...\ran [C] = \lan p_1|...\ran [p\cap C] =
          ( p \cdot \lan p_1 |...\ran ) [C] \]
which means that a {\em vertical}
rational curve in $X_G$ which has a common point with the preimage
$\pi ^{-1}(p)$ of a finite codimension cycle $p\subset BG$ in the base,
is entirely contained in this preimage.
\een

\bigskip
Similarly to ordinary quantum cohomology, quantum equivariant intersection
indices have a few other interpretations.

\subsection{Intersections in `moduli spaces'}

Consider the product $X^N$ of $N$ copies of $X$ as a $G$-manifold provided with
the diagonal $G$-action. The homotopic quotient $X^N_G$ has $N$ canonical
projections $X^N_G\to X_G$ compatible with the projections $X^N_G\to BG$,
$X_G\to BG$ to the classifying space. Let $p_1,...,p_N$ be equivariant
cohomology classes of $X$. One may think of them as represented by finite
codimension cycles in $X_G$, one in each of $N$ copies. Pulled back to $X^N_G$
they define $N$ equivariant cohomology classes of $X^N$ which we denote
$p_1,...,p_N$ too.

Let $\calm _d$ denote the space of parametrized rational curves
$\gf : \CC P^1 \to X$ of certain degree ($=$ homology class) $d$.
Evaluation map $\calm _d\to X^N, \ \gf \mapsto \gf (x_1),...,\gf _N(x_N)$
at $N$ generic points in $\CC P^1$ defines a $G$-invariant complex
subvariety in $X^N$. Its fundamental cycle $\bar{\calm }_d$ determines an
equivariant cohomology class of $X^N$: it is Poincare-dual to
\[ EG\times _G \bar{\calm }_d \subset EG\times _G X^N .\]
%
We denote this equivariant class $[\calm _d]$.

One defines the quantum equivariant intersection index using classical
equivariant indices in $H^*_G(X^N)$ as
\[ \lan p_1 |... |p_N \ran = \sum _d \lan p_1,...,p_N, [\calm _d] \ran q^d .\]
It is easy to see what is the meaning of the RHS, evaluated at a cycle
$C\subset BG$: it counts the
numbers of discrete rational maps $\gf $ to the fibers of the bundle
$\pi ^{-1} (C) \to C$ such that $\gf (x_i)$ is in the cycle representing $p_i$
in $X_G$. The maps are ``weighted'' by the factors $q^d$ and are counted with
the signs prescribed by (co)orientations of the cycles. In particular, this
construction (being at least morally equivalent to the first one) specifies
how the signs $\pm $ in the previous definition should be chosen.

\subsection{Integrals in loop spaces}

The quantum intersection indices defined by means of evaluation maps are
(expected to be) independent on the choice of evaluation points $x_1,...,x_N$
on the projective line provided that the points are generic (and in particular
distinct). Therefore one can replace $\lan p_1,...,p_N , [\calm _d (x)] \ran$
by its average value
\[ \int _{T^N} \lan p_1,...,p_N,[\calm _d(x)] \ran dx_1 ... dx_N \]
where $T^N$ is a torus in the configuration space $(\CC P^1)^N$ of $N$ points
$x=(x_1,...,x_N)$, namely the product of $N$ standard unit circles in
$\CC P^1=\CC \cap \infty $ (notice that $T^N$ is dense in Zarissky topology
on $(\CC P^1)^N$). This formula allows us to interpret the intersection
indices as some integrals of differential forms on loop spaces.

Suppose that the classifying space $BG$ is chosen in the form of
infinite-dimensional manifold and that the equivariant cohomology classes
$p_1,...,p_N$ are represented by closed differential forms on $X_G$.
Such a differential form determines a differential form of the same degree
on the space of free loops in $X_G$. Namely, if $t\mapsto \gc (t)$ is
a loop, the average $\oint p_{t} dt$ is an exterior form on the space of vector
fields along the loop, and thus $P=\oint p _dt$ is a differential form on the
loop space, closed if $p$ is closed on $X_G$.

Furthermore, we interpret a (vertical) rational curve $\gf : \CC P^1 \to X_G$
as an ``algebraic loop'' restricting the map $\gf $ to the unit circle
$T\subset \CC - 0 \subset \CC P^1 $. Now on we may think of the spaces
$\calm _d$ of rational maps, as well as of the spaces $\calm _d [C]$
of such vertical rational maps to the fibers of the bundle $X_G\to BG$ over
a given cycle $C\subset BG$,  as subsets (chains, cycles) in the loop space.

The above integral over the torus immediately turns into the integral in the
loop space,
\[ \lan p_1,...,p_N,[\calm _d] \ran [C] =
    \int _{\calm _d [C]} P_1\wedge ... \wedge P_N .\]
As usual, this formula assumes that the integral equals zero unless the
total degree $M$ of the wedge product equals the dimension $c(d)+D+K$ of
the chain $\calm _d [C]$.

We will make use of this construction in the special case when the equivariant
cohomology algebra $H^*_G(X)$ is generated (as algebra) by the classes of
degree $2$ --- that is of degree $1$ taking into account our convention that
all the dimensions and degrees are complex. Let $p_1,...,p_n$ now denote
a set of such generators, \this a basis in $H^2_G(X)$. We prefer to think of
$p_i$ as of closed differential $2$-forms on the infinite-dimensional
manifold $X_G$, or even as of symplectic (or Kahler) forms, taking into
account our assumptions about $X$ and the fact that classifying spaces of
compact Lie groups have Kahler models. Denote
\[ P(z)=z_1 P_1 + ... + z_n P_n \]
a general linear combination of the differential (symplectic, Kahler) $2$-forms
$P_i$ on the loop space of $X_G$ corresponding to the forms $p_i$ on $X_G$.
Let us define the {\em generating volume function} $V\in H^*_G(pt,
\ZZ [[z,q]])$
--- a formal series in $q$ and $z$ with coefficients in the ring of
characteristic classes, such that the value of $V$ on a homology class
represented by the cycle $C\subset BG$ is equal to the weighted oriented volume
\[ V|_{[C]}=\sum _d q^d \int _{\calm _d [C]} \exp{ (P(z))} \]
of the space $\calm [C]=\cup _d q^d \calm _d [C] $ of vertical rational curves
over $C$. Here $\exp{(P)}$ stands for
\[ \sum _{k=0}^{\infty } \frac{1}{k!} P\wedge ... \wedge P \ (k\ \text{times})
\]
so that the integral $\int _{\calm } \exp{(P)} $ really represents the
symplectic $k$-dimensional volume of a $k$-cycle $\calm $ if the form $P$ is
symplectic (we should notice
however that orientation of $C$ contributes the sign of the ``volume'').

The generating volume function has not so many non-zero terms as one could
think: due to dimension reasons it is {\em weighted-homogeneous of degree $-D$}
when the degrees of the variables are assigned as
\[ \deg q^d = c(d), \ \deg z_i = -1 \]
and characteristic classes from $H^*_G(pt)$ have their natural degrees.

\bigskip
One of applications of this function describes quantum intersection indices
of the generators $p_i$:
\[ \lan p_{i_1}|...|p_{i_N} \ran =
\frac{\p ^N}{\p z_{i_1} ... \p z_{i_N}} |_{z=0} \ V (z) \]
(it is just the property of the exponential series).

\bigskip
Another property of the volume generating functions, that we are going to
exploit, is their simple behavior under {\em product, restriction} and
{\em induction} operations.

\bigskip
{\bf Product.} Let $X'$, $X''$ be compact Kahler $G'$- and $G''$-spaces
respectively, and $V'(z',q')\in H^*_{G'}(pt)$, $V''(z'',q'')\in H^*_{G''}(pt)$
be the corresponding generating volume functions. Then the generating
volume function $V$ for the $G'\times G''$-space $X'\times X''$ is
\[ V((z',z''),(q',q''))=V'(z',q')V''(z'',q'') .\]

Indeed, the homotopic quotient of $X'\times X''$ is the product of $X'_{G'}$
and $X''_{G''}$ fibered over the product $BG'\times BG''$ of classifying
spaces. A holomorphic map to $X'\times X''$ is a pair of holomorphic maps
to $X'$ and $X''$ respectively and hence the chain $\calm _{d',d''}$ factors:
\[ \calm _{d',d''}[C'\times C'']=\calm _{d'}[C']\times \calm _{d''}[C''] .\]
Its volume with respect to $P(z)=P'(z')\oplus P''(z'')$ is the product of
corresponding volumes and therefore

\[ \sum _{(d',d'')} (q')^{d'}(q'')^{d''}
      \int _{\calm _{d'}[C'] \times \calm {_d''}[C'']}
      \exp (P(z)) = \]
\[    [\sum _{d'} (q')^{d'} \int _{\calm _{d'}[C']} \exp (P'(z'))]  \cdot
    [\sum _{d''} (q'')^{d''} \int _{\calm {_d''}[C'']} \exp (P''(z''))] .\]

\bigskip
{\bf Restriction.} Let $X$ be a compact Kahler $G$-space and $G'\subset G$ be
a Lie subgroup. Considering $X$ as a $G'$-space, we obtain an $X$-bundle
$X_{G'}\to BG'$ (induced, as a bundle, from $X_G\to BG$ by means of the
natural map $\pi : BG' \to BG$ of classifying spaces) and the
corresponding map of total spaces $\gz : X_{G'}\to X_G$ with the fiber $G/G'$.
Then for the generating volume functions $V(z,q)$ and $V'(z',q)$ we have
\[   V'(\gz ^*(z),q) = \pi ^* V(z,q) .\]

Indeed, for a cycle $C'\subset BG'$ the bundle $\calm _d [C'] \to C'$ is
induced by $\pi $ from $\calm _d [\pi _*C']\to \pi (C')$ and therefore
\[ \int _{\calm _d[C']} \exp (\gz ^*(P(z)) =
\int _{\calm _d[\pi _*C']} \exp (P(z)) .\]

In particular, if $G'$ is trivial so that $\pi$ is $ EG\to BG$ and
$X_{G'} = EG\times X$, then the homomorphism $\gz ^*:H^2(X_G)\to H^2(X),\
z \mapsto z',$ is onto, and the generating volume function $V'(z',q)$
coincides with the non-equivariant one and can be computed from $V(z,q)$
as its reduction $H^*_G(pt)\to \ZZ$ modulo $G$-characteristic classes of
positive degree.

This implies that non-equivariant quantum intersection indices
$\lan p_{i_1}|...|p_{i_N} \ran $ are obtained by such a reduction
from the corresponding quantum equivariant intersection indices.

\bigskip
{\bf Induction.} Let $G'\subset G$ be a subgroup with a simply-connected
compact Kahler quotient
$G/G'$, and $Y$ be a compact Kahler $G'$-space. We construct a compact Kahler
$G$-space $X=G\times _{G'} Y$ and call it {\em induced} from $Y$ (like induced
representations). In fact $X$ is fibered over $G/G'$ with the fiber $Y$.
The homotopic quotient spaces of $X$ and $Y$ coincide:
\[ X_G=EG\times _G (G\times _{G'} Y) = EG\times _{G'} Y = Y_{G'} ,\]
and thus their equivariant cohomology is the same, but the module structure
in $H^*_G(X)$ is induced from the module structure in $H^*_{G'}(Y)$ by the
natural map $BG'\to BG$.

Let $p''$ be a basis of non-negative classes in $H^2(G/G')$ lifted to $X$, and
$p=(p',p'')$ be its extension to such a basis in $H^2(X)$. Encoding the
homology class of a rational curve in $X$ by the string
$(d',d'')=(d_1,...,d_k)$
of its degrees with respect to the dual basis in $H_2(X)$, we find that
the curves vertical in the bundle $X\to G/G'$ have  $d''=0$ and
{\em vice versa}.

This means that the quantum deformation ring $\ZZ [q']$ for $Y$ can be
considered as a {\em quotient} of the corresponding ring for $X$:
\[ \ZZ [q'] = \ZZ [q',q'']/(q'') .\]

\bigskip
{\em Remark.} This identification may seem confusing, since the group algebra
$\CC [q'{}^{\pm 1}]$ is a {\em subalgebra} in $\CC [q^{\pm 1}]$. In fact,
replacing the algebra $\CC [q^{\pm 1}]$ of functions on the torus by the
polynomial algebra $\CC [q]$ defines, in geometrical terms,  partial
compactification of the torus to $\CC ^k$. Our description of $\CC [q']$ as
a quotient corresponds to the embedding of such a compactified torus
$\CC ^{k'}$ for $Y$ into the ``boundary'' $\CC ^k - (\CC - 0)^k$ of the torus
for $X$.

\bigskip
Denote $V'(z,q')$ and $V(z,q)$ the generating volume functions for quantum
equivariant cohomology of $Y$ and $X$ respectively. Then
\[ V(z,(q',0)) = \text{Direct image} \ V'(z,q') \]
where the direct image operation refers to the bundle $\pi: BG' \to BG$.

\bigskip
Indeed, when we evaluate $V(z,q)$ on some cycle $C\subset BG$ at $q''=0$ we
simply calculate weighted volume of the space of vertical algebraic loops in
$X_G$ over $C$ but throw away contributions of all rational curves with
$d''\neq 0$. But a rational curve in $X$ with $d''=0$ projects to $G/G'$ to a
point. This means that the LHS actually computes weighted volume of the space
of vertical algebraic loops in $Y_{G'}\to BG'\to BG$ over the preimage
$C'=\pi ^{-1} C$. Therefore
\[ V(z,(q',0)) [C] = V'(z,q')[C'] = [\text{Direct image}\ V'(z,q')][C] \]
by the very definition of the direct image operation.

\subsection{Equivariant Floer homology}

We briefly discuss here quantum equivariant cohomology from the point of view
of Morse-Floer theory on loop spaces. This discussion is supposed to motivate
our conjecture that the general properties expected from quantum cohomology can
be naturally generalized to the equivariant case.

\bigskip
Let $X$, as above, be a compact simply-connected Kahler manifold provided with
a holomorphic action of the complexified compact Lie group $G_{\CC }$ and
and with a $G$-invariant Kahler form.
The group $G_{\CC }$ also acts by holomorphic transformations on the loop space
$LX$ and its universal covering.
Since the action functional $\cala $ on the covering is
$G$-invariant one can try to construct the equivariant Floer (co)homology
$FH^*_G(X)$ by means of equivariant Morse-Witten theory for $\cala $.

Usually one defines an equivariant Morse chain complex using finite-dimensional
approximations $EG_N \to BG_N$ of the universal $G$-bundle. For example, if $G$
is the unitary group $U_n$ one can choose the complex Grassmann manifold
$Gr(N,n)$ on the role of $BG$ and the corresponding Stiefel manifold on the
role
of $EG_N$. Mimicking this approach, we can extend the functional $\cala $ to
the
space $EG_N\times LX^{~}$ in the trivial manner and thus construct a functional
$\cala _N$ on the manifold $L_N=EG_N\times _G LX^{~}$ approximating the
homotopic
quotient space $(LX^{~})_G$. Now we can apply Floer's semi-infinite Morse
theory
to the functionals $\cala _N$.
Notice that the homotopic quotient $(LX)_G$ is
nothing but the space of vertical loops in the bundle $X_G\to BG$, and $L_N$ is
simply its restriction to $BG_N\subset BG$.

Taking care of the riemannian metric, add a $G$-invariant riemannian metric on
$EG_N$ as a direct summand to the Kahler $G$-invariant metric on $LX^{~}$
induced
from that on $X$. Then the gradient vector field of $\cala $ on
$EG_N\times LX^{~}$ is tangent to the second factor and is invariant with
respect
to the diagonal action of $G$. This means that the gradient vector field of
$\cala _N$ relative to the factor-metric on $L_N$ is just the projection of
that $G$-invariant field, and the corresponding gradient flow consists in
{\em fiberwise} analytic continuation of vertical loops in the $X$-bundle
$L_N\to BG_N$. In particular, Floer cohomology of $L_N$ will carry a module
structure over the ordinary cohomology algebra of $BG_N$.

Notice that the $G$-action on $LX^{~}$ commutes with both the circle action
($=$ reparametrization of loops) and the action of the covering transformation
group $\ZZ ^k$ (so that both actions survive on $L_N$). The first implies
that $\cala _N$ is a perfect Morse-Bott function on $L_N$ (see [G],[G1]). The
second describes the action of the group ring $\ZZ [q^{\pm 1}]$ on the Floer
cohomology of $L_N$, which is therefore additively isomorphic to the cohomology
$H^*((X_G)_N, \ZZ [q^{\pm 1}])$ of the critical point set.

Passing to the limit $N\to \infty $, we conclude that $G$-equivariant Floer
cohomology $FH^*_G(X)$ of $LX^{~}$ should be a
$H^*_G(pt, \ZZ [q^{\pm 1}])$-module
canonically isomorphic to the equivariant cohomology of $X$ with coefficients
in the group ring $\ZZ [q^{\pm 1}]$.

\bigskip
A multiplicative structure in equivariant quantum cohomology of $LX$ can be
defined by means of the evaluation map at the point $1\in S^1$:
\[ LX^{~}\to (LX^{~}\times X), \ \ (\gc: S^1\to X) \mapsto (\gc , \gc (1)) .\]
This map is $G$-equivariant and induces an action of equivariant cohomology
classes of $X$ by module endomorphisms on equivariant Floer cohomology
$FH^*_G(X)$ of the loop space $LX^{~}$. Using our explicit description of the
gradient flow on $(LX)_G$ as fiberwise analytic continuation of loops, one
can compute this action in terms of vertical holomorphic curves and quantum
equivariant intersection indices $\lan |...| \ran $ introduced in $3.3$.
Namely the action of $p\in H^*_G(X)$ on $a\in FH^*_G(X)$ satisfies
\[  \lan p*a, b \ran = \lan a | p | b \ran \]
for any $b\in FH^*_G(X)$ where the pairing on the LHS is the classical
equivariant intersection index on $H^*_G(X, \ZZ [q^{\pm 1}])$ with values
in $H^*_G(pt, \ZZ [q])$.

The multiple quantum equivariant intersection indices
$\lan a|p_1|...|p_r|b\ran $ can be expressed in a
similar manner in terms of evaluation maps $LX \to LX \times X^r$ at $r$
distinct points $x_1,...,x_r$ on the circle $S^1$. We conjecture that they
satisfy the ``principal axiom'' of Topological Field Theory (see $2.2$).
This conjecture implies that the multiple intersection indices represent
matrix elements of compositions of the endomorphisms corresponding to
$p_1,...,p_r \in H^*_G(X)$. Finally, if one defines quantum equivariant
cohomology of $QH^*(X)$ as the algebra generated by these endomorphisms and
operators of multiplication by $q$, then our conjecture means that this
algebra
\begin{itemize}
\item is additively isomorphic to $H^*_G(X, \ZZ [q])$ (or may be ``$[[q]]$''),
\item provides a ``quantum'' deformation of the classical equivariant
cohomology
algebra $H^*_G(X)$,
\item inherits the module structure over $H^*_G(pt)\otimes \ZZ [q]$, and
\item allows to express the multiple pairings through quantum multiplication
and the classical direct image functional:
\[ \lan p_1|...|p_r\ran = \lan p_1...p_r , [1] \ran .\]
\end{itemize}

\bigskip
It is difficult to say now whether a rigorous justification of these hypotheses
should be even more sophisticated than in the non-equivariant case.
One one hand, general position arguments should require introducing almost
complex structures on $X$ which are not $G$-invariant. The most natural way to
handle this problem --- by considering the space $\calj $ of all almost complex
structures and constructing $G$-equivariant Floer cohomology of
$LX\times \calj $ --- involves one more ``infinity'' and seems to raise the
level of technical difficulty. On the other hand, the finite-dimensional
approximations $BG_N$ of classifying spaces have Kahler models, and quantum
equivariant cohomology of $X$ seem to be expressible in terms of
non-equivariant quantum cohomology of the approximations $(X_G)_N\to BG_N$:
it suffices to ``throw away'' contributions of non-vertical rational curves
in $(X_G)_N$, \this put some of `$q$'s equal zero. This approach can possibly
reduce the problem back to the axioms of non-equivariant Topological Field
Theory.

We are not ready to discuss further this problem here. We also leave for the
reader to think out the parallel construction of equivariant quantum
multiplication which is based on composition of loops.

\subsection{Characteristic classes as Casimir functions}

Here we interpret the quantum equivariant cohomology algebra $QH^*_G(X)$ as
the algebra of functions on some lagrangian variety in the assumption that
the ordinary cohomology algebra $H^*(X)$ of the simply-connected Kahler
manifold
$X$ is generated by non-negative $(1,1)$-classes $p_1,...,p_k$ (in notations of
$2.3$). This assumption along with the spectral sequence of the $X$-bundle
$X_G\to BG$ implies that the equivariant cohomology algebra $H^*_G(X)$ is
additively isomorphic to $H^*(BG)\otimes H^*(X)$ and is generated, as an
$H^*_G(pt)$-algebra, by $k$ elements representing $1\otimes p_i$ which we will
denote $p_1,...,p_k$ again.

Its quantum deformation $QH^*_G(X)$ has been
defined in $3.7$ by means of the identity
\[ \lan a*b, c \ran = \lan a|b|c \ran .\]
Considered as $H^*_G(pt)$-algebra, it is generated by
$(p_1,...,p_k,q_1,...,q_k)$ and is therefore isomorphic to the quotient of the
polynomial algebra $ H^*_G(pt)[p,q]] $ by some ideal of relations.

Passing to complex coefficients and introducing temporary notations
$c_i$, $i=1,..., r$,
for generators of the polynomial algebra $H^*(pt,\CC )=\CC [c] $
of $G$-characteristic classes, we interpret the quantum equivariant cohomology
algebra $QH^*_G(X,\CC )$ as the algebra of regular functions on a
(quasi-homogeneous) subvariety $\call $ determined by the ideal of relations
$\cali $ in the complex space with coordinates
\[ (p_1,...,p_k,q_1,...,q_k,c_1,...,c_r) .\]

This complex space has a natural Poisson structure
\[  q_1 \frac{\p }{\p q_1} \w \frac{\p }{\p p_1} +
  ... + q_k \frac{\p }{\p q_k} \w \frac{\p }{\p p_k} \]
due to the constant coefficient pairing between $H^2(X)=H^2(X_G)/H^2(BG)$ and
$\ZZ ^k=H_2(X)\subset H_2(X_G)$ (we assume of course that the basis in the
lattice $\ZZ ^k$ is dual to the basis $(p_1,...,p_k)$ in $H^2(X)$).

\bigskip
We observe that the characteristic classes $c_i$ play the role of Casimir
functions of such a Poisson structure and claim that the characteristic variety
$\call $ is lagrangian in the sense of Poisson geometry, \this its
intersections
with the symplectic leaves $\vec{c}=const$, $q_1...q_k\neq 0$ are lagrangian at
their regular points.

\bigskip
Similarly to the non-equivariant case $2.3$, this statement is based on the
properties of the matrix-valued differential $1$-form
$A=\sum (p_i*)(dq_i)/q_i$ to satisfy $dA=0, A\w A=0$, but now the
Casimir functions $c_i$ are treated by the differential $d$ and by the
operators
$p_i*$ as constants. Mimicking $2.3$, we introduce a $\CC [[c,q]]$-valued
bilinear form on $H^*_G(X, \CC[[q]])$ by the formula
\[ S_{a,b} |_{[C]} = \sum _{\begin{array}{c}
   \text{degree}\, d \ \text{isolated vertical rational curves}  \\
      \text{in}\ X_G\to BG \ \text{restricted to}\ C\subset BG  \\
      \text{ with two marked points in}\, a \, \text{and}\, b \end{array} }
\pm q^d \]
which evaluates the bilinear form of two finite codimension cycles
$a,b \subset X_G$ on a finite-dimensional cycle $C\subset BG$.

Thinking of $c_i$ as of the preimage in $X_G$ of a finite-codimension cycle
in $BG$ we immediately conclude that $S$ is $\CC [c]$-bilinear:
\[ S_{c_ia,b} |_{[C]} = S_{a,b} |_{[c_i\cap C]} = (c_i S_{a,b})|_{[C]} .\]
Thinking of $p_i$ as a complex hypersurface in $X_G$ we find, as in $2.3$, that
\[ S_{a,b} + \sum \lan a, p_i, b \ran \log (q_i) \]
is a potential for the $(a,b)$-matrix element of the $1$-form $A$:
\[ q_i\frac{\p }{\p q_i} S_{a,b} = \lan a|p_i |b\ran - \lan a,p_i,b \ran  .\]
This is equivalent to $dA=0$ and together with commutativity $A\w A=0$ and the
lemma in $2.3$ implies that each non-singular branch of
$\call \cap \{ \vec{c} = const \} $ over the torus with coordinates $q$ is
lagrangian in the cotangent bundle of this torus ($=$ the symplectic leaf with
coordinates $p,q$).

\section{Computation of $QH^*_{U_n}(F_n)$}

In this section, we compute quantum (equivariant) cohomology of flag manifolds.
The results here are mathematically rigorous {\em corollaries} of the following
{\em conjectures} about general properties of quantum cohomology of Kahler
manifolds:
\begin{itemize}
\item Quantum equivariant cohomology is a skew-commutative associative algebra
over the characteristic class ring;
\item It is a weighted-homogeneous $q$-deformation of the classical equivariant
cohomology;
\item Equivariant generating volume functions satisfy the product, restriction
and induction properties from $3.6$.
\end{itemize}

\subsection{Root systems}

The structure of the $2$-nd (co)homology lattice of flag manifolds can be
understood better in terms of root systems. The flag manifold $F_n$ is the
space $G_{\CC }/B$ of all Borel subalgebras in
$\frak{g} _{\CC }=\frak{sl} _n(\CC )$. Therefore its tangent bundle splits
canonically into the direct sum of line bundles $\oplus _{\ga } L_{\ga }$
indexed by {\em positive roots} $\ga $ of the root system $A_{n-1}$. Recall
that this root system can be described as the set of linear functions
$x_i-x_j$ on the lattice $\ZZ ^n$ with coordinates $x_1,...,x_n$, and
the positive roots are those with $i<j$. The $n-1$-dimensional lattice spanned
by the roots can be identified with a finite index sublattice in the $2$-nd
cohomology group $H^2(F_n)$ by the map
    \[ \text{a line bundle} \mapsto \text{its $1$-st Chern class} .\]
Therefore the $1$-st Chern class of the flag manifold is represented by
the total sum $2\gr $ of positive roots.
   According to Borel-Weil theory, finite-dimensional representations of $SU_n$
can be realized in spaces of holomorphic sections of non-negative line bundles
over $F_n$ and correspond in a $1-1$ fashion to their $1$-st Chern classes.
This
theory implies that the Kahler cone of $F_n$ is the Weyl chamber spanned by the
$1$-st Chern classes $p_1,...,p_{n-1}$ of the {\em fundamental} line bundles
$\det ^*\gL ^i\CC ^n, i=1,...,n-1$, called --- in terms of the root system ---
{\em fundamental weights}.

The fundamental weights $p_i=x_1+...+x_i, i=1,...,n-1$, form a basis in the
lattice $H^2(F_n)$. The vectors $\ga _1,...,\ga _{n-1}$ of the dual basis  and
their non-negative integer combinations represent, in the homology
group $H_2(F_n)$,  classes of {\em holomorphic} curves in $F_n$. Identifying
the space $H^*(F_n, \QQ)$ with its dual by means of the Weyl-invariant inner
product (the Cartan matrix is its matrix in the basis of fundamental weights)
we find that $(\ga _1,...,\ga _{n-1})$ becomes the basis of {\em simple} roots
$\ga _i=x_i-x_{i+1}$ under this identification.

Now the famous identity
\[ \sum _{\ga >0} \ga = 2\gr = 2(p_1 +...+p_{n-1}) \]
along with $\lan p_i, \ga_j \ran = \gd _{ij} $ means that in our representation
of classes $\sum d_i\ga _i$ of rational curves by monomials
$q_1^{d_1}...q_{n-1}^{d_{n-1}}$ the degrees of the variables $q_i$ are
\[ \deg q_i =  c(\ga _i) = \lan 2\gr , \ga _i \ran = 2 .\]

\subsection{Auxiliary bundle}

According to general theory,
\[ QH^*_{U_N}(F_n)=\ZZ[u_1,...,u_n,q_1,...,q_{n-1}, c_1,..., c_n]/I_{U_n} \]
where the ideal $I_{U_n}$ is generated by some quasi-homogeneous
$q$-deformation
of the relations
\[ c_i=\gs _i(u), i=1,...,n, \ \ \deg u_i=1, \deg c_i= i, \deg q_i=2 \]
which can be written (using a formal variable $\gl $ of degree $1$) as a
single quasi-homogeneous identity of degree $n$:
\[  (u_1+\gl) ... (u_n+\gl) =
\gl ^{n} + \gs_1 \gl ^{n-1} + ... + \gs _{n} .\]
We find this deformation by induction on $n=2,3,4... $, based on the
following obvious

\bigskip
{\bf Lemma 1.} {\em For $n>2$, suppose that a quasi-homogeneous relation  of
the form
\[  (u_0+\gl) ... (u_n+\gl) -
[\gl ^{n} + \gs_1 \gl ^{n-1} + ... + \gs _{n}]
= O(q_1, ... , q_{n-1}) [\gl ,q,u,\gs ] \]

is satisfied in quantum equivariant cohomology algebra of the flag manifold
$F_n$ modulo $q_i$ for each $i=1, ... , n-1$.
Then this relation holds identically (\this for all $q$).}

\bigskip
{\em Proof.} Indeed, since the LHS of the
relation in question is homogeneous of degree $n$, the hypothesis of Lemma $1$
means that the difference $LHS - RHS$
is divisible by $q_1...q_{n-1}$. But $\deg q_i = 2$ and
\[ \deg q_1...q_{n-1} =2n-2 > n \ \text{for} \ n > 2 .\]
This implies that $LHS-RHS=0$.

{\em Remark.} This lemma is the only place in our proof where we use some
specificity of the group $U_n$. It also holds for flag manifolds of series $C$
and
$D$ but fails for other compact simple Lie groups. For their flag manifolds
one can easily give a hypothetical description of the quantum equivariant
cohomology algebras in terms of generalized Toda lattices, but a proof
should involve some additional argument.

\bigskip
Our inductive step will make use of the following construction.
Consider the subgroup $G'=U_m\times U_{n-m} \subset U_n=G$ and the $G'$-space
$Y=F_m\times F_{n-m}$. The induced $G$-space (in the sense of $3.6$) is
nothing but the flag manifold $F_n$. Its fibration over $G/G'=Gr(n,m)$
sends a flag in $\CC ^n$ to its $m$-dimensional component.

Let $V_m$ denote generating volume function for quantum equivariant cohomology
of $F_m$.

\bigskip
{\bf Lemma 2.}
\[ V_n(z,q,c) |_{q_m=0} =
   \text{\em Direct image}\ [V_m(z',q',c')\cdot V_{n-m}(z'',q'',c'')] \]
{\em where

$z=(z_1,...,z_n)$ are coordinates on $H^2((F_n)_G)$ with respect to the
basis $u_1,...,u_n$ (see $3.2$), $z'=(z_1,...,z_m), z''=(z_{m+1},...,z_n)$,

$q = (q', q_m, q'') = (q_1,...,q_m, ... ,q_{n-1})$,

$c$, $c'$ and $c''$ are Chern classes of $U_n$, $U_m$ and $U_{n-m}$
respectively, and ``{\em Direct image}'' refers to the direct image operation
$\ZZ[c',c'']=H^*(BG') \to H^*(G)=\ZZ [c]$ for the bundle $BG'\to BG$ with the
fiber $Gr(n,m)$ (see $3.3$).}

{\em Proof.} It is a straightforward corollary of the product and induction
formulas: factorization $(F_n)_{U_n}=(F_m)_{U_m}\times (F_{n-m})_{U_{n-m}}$
identifies the basis $(u_1,...,u_n)$ in the $2$-nd equivariant cohomology
of the product
with the union $(u'_1,...,u'_m, u''_1,...,u''_{n-m})$ of such basises of
factors
since both are the standard generator sets in the cohomology of
$(\CC P^{\infty })^n$, and $p_m\in H^2(F_n)$ is represented by the $1$-st
Chern class of the determinant line bundle over $Gr(n,m)$ and therefore
the vertical rational curves in $F_n\to Gr(n,m)$ are exactly those with
$d_m=0$.

\subsection{Theorem 2 implies Theorem 1}

Indeed, according to the {\em restriction} property of equivariant generating
volume functions (applied to the trivial subgroup in $U_n$), if a relation
\[ R(\p /\p z, q, c) V(z,q,c) = 0 \]
is satisfied, then $R(\p /\p z,q,0)$ annihilates the non-equivariant generating
volume function \newline
$V(z,q,0)$ and thus the relation $R(u,q,0)=0$ holds in
$QH^*(F_n)$. This proves

\bigskip
{\bf Lemma 3.}
\[ QH^*(F_n)=QH^*_{U_n}(F_n) /(c_1,...,c_n) .\]

\subsection{Equivariant quantum cohomology of $\CC P^1$}

{\bf Lemma 4.}
\[ QH^*_G(\CC P^1) =
\ZZ [u_1,u_2,q,c_1, c_2]/(u_1+u_2=c_1, u_1 u_2 + q =c_2) \]

\bigskip
{\em Proof.} Quantum equivariant cohomology of the projective line $F_2$
is isomorphic to the quotient algebra of $\ZZ [u_1,u_2,q,c_1,c_2]$ by the
ideal generated by quantum deformations of the
relations $u_1+u_2=c_1, u_1 u_2 = c_2$ in the classical equivariant cohomology.
These deformations can be taken quasi-homogeneous and since $\deg q = 2$, the
only possible deformation should replace the RHS in $u_1u_2-c_2 = 0$
with a scalar multiple of $q$.

In order to determine the scalar it suffices to reduce the relations modulo
$(c_1,c_2)$, \this to compare, by Lemma 3, with relations in the ordinary,
non-equivariant quantum cohomology of $F_2=\CC P^1$. Then $u_2=-u_1$
represents the $1$-st Chern class of the ``hyperplane'' bundle
over $\CC P^1$, \this simply a point. Since the relation $u_2^2=q$ holds in
the quantum cohomology of $\CC P^1$ (see $2.7$), the scalar coefficient we are
looking for equals $1$.

\subsection{Step of induction}

Denote
\[ D_n(u,q,\gl )=\det (A_{n-1} +\gl ) \]
the characteristic polynomial of the $n\times n$-matrix
with $u_1,...,u_n$ on the diagonal, $q_1,...,q_{n-1}$ right above
and $-1,...,-1$ right under the diagonal.

{\bf Lemma 5.} {\em Suppose that the relation
\[ D_k(u,q,\gl ) = \gl ^k + c_1\gl ^{n-1} + ... + c_k \]
is satisfied identically in $\gl $ in the equivariant quantum
cohomology of flag manifolds $F_k$ for all $k<n$.
Then the relation with $k=n$ is also satisfied modulo $q_m$ for every
$m=1,...,n-1$.}

{\em Proof.} First of all,
notice that $D_n |_{q_m=0} = D_m(u',q',\gl ) D_{n-m}(u'',q'',\gl )$
where $(u',u'')=u, (q',0,q'')=q$.

Denote
\[ \gS _n = \gl ^n +c_1\gl ^{n-1}+...+c_n= (x_1+\gl )...(x_n+\gl ) \]
the RHS of the above relation with the Chern classes $c_1,...,c_n$ written
for convenience as elementary symmetric functions of the formal variables
$x_1,...,x_n$.
The conclusion of Lemma $5$ means that
\[ [D_n(\p /\p z,q,\gl ) - \gS _n(x,\gl )]\, V_n(z,q,\gs (x))\ |_{q_m=0} = 0
.\]

It is the same as
\[ [D_m(\frac{\p}{\p z'},q',\gl ) D_{n-m}(\frac{\p}{\p z''}, q'',\gl )
- \gS _n(x,\gl )]
 \ [V_n((z',z''),(q),\gs (x))|_{q_m=0}] = 0 .\]

By Lemma $2$, the function $V_n|_{q_m=0}$ in the last formula can be
replaced with the {\em Direct image} of
\[ V_m(z',q',\gs (x')) \cdot V_{n-m}(z'',q'',\gs (x'')) ,\]
explicitly described in $3.3$.

Since the derivations in $D_m D_{n-m}$ are with respect to $z',z''$
which are not involved into permutations in the operation $Alt$, and the
variables $x',x''$ which are involved do not show up in coefficients of the
operators $D_m, D_{n-m}$, the {\em Direct image} operation commutes with
our differential operator.

Applying the inductive assumption we find that the conclusion of the
proposition is equivalent to the identity
\[ \gS _n(x,\gl )  \text{Direct image}\ [V_m (x') V_{n-m} (x'')] = \]
\[  =  \text{Direct image} \ [ (\gS _m(x',\gl )  V_m(x'))
    (\gS _{n-m}(x'',\gl ) V_{n-m}(x''))] .\]

But
\[ \gS _m(x',\gl ) \gS _{n-m}(x'',\gl ) =  (x_1+\gl )...(x_n+\gl )
= \gS _n(x,\gl ) \]
is totally symmetric in $(x_1,...,x_n)\,$!.

Since multiplication by a symmetric function commutes with the alternation
operation, we conclude that the required identity does hold.

\bigskip
Combining Lemma $5$ with Lemma $1$ completes the proof of Theorem $2$ from
Introduction.

\subsection{Volume functions}

We have found the relations in quantum cohomology of flag manifolds using
general properties of generating volume function. Now we compute the quantum
volume functions using our knowledge of the relations and of the classical
volume functions.

\bigskip
{\bf Proposition.}{\em  The quantum equivariant generating volume function
$V_n(z,q,c)$ of the flag manifold $F_n$ equals
\[ W_n  = \frac{1}{(2\pi i)^n} \, \int
\frac{\exp (z,u) du_1 \w ... \w du_n}{(\gS _1(u,q)-c_1)...(\gS _n(u,q)-c_n)} \]
where $\gS _i(u,q)$ are the quantum deformations of elementary symmetric
functions \this the coefficients of the polynomial $\det (A_{n-1}+\gl )$.}

\bigskip
{\em Proof.}
By the {\em deformation} property and Proposition in $3.2$, the formula holds
for $q=0$. We will prove the formula using
the homogeneity property $\deg V_n = -\dim F_n $ (where $\deg z_i=-1,
\deg q_i=2, \deg c_i=i$) and the differential equations
\[ \gS _i (\p /\p z, q) V_n(z,q,c) = c_i V_n(z,q,c), \ i=1,...,n .\]

\bigskip
First of all, the function $W_n$
does satisfy the homogeneity condition  and the differential equations (due to
the famous property of residues).

Due to another property of residues (see [GH]) $W_n$ is an analytic function of
its variables and can be expanded into a power series ($V_n$ is a formal series
by definition). Represent the difference $V_n-W_n$ as a sum
$\sum _{d\geq 0,l\geq 0} R_{d,l}(z) q^d c^l $.
The coefficient $R_{d,l}$ is a homogeneous polynomial in $z$ of degree (in the
usual sense)
$\dim F_n + \sum 2d_i+ \sum jl_j$ and $R_{0,0}=0$ since $V_n$ coincides with
$W_n$ at $q=0$.

Let us pick $R$ as the coefficient of minimal degree among non-zero $R_{d,l}$.
The differential equations for $V-W$ mean that
\[ \gs _i (\p /\p z) R(z) = \text{some operators applied to}\ R_{d,l}
                            \ \text{with smaller}\ d,l \]
and hence that $\gs _i(\p /\p z) R(z)=0, i=1,...,n$, since all those
$R_{d,l}$ are zeroes. Now the following lemma completes the proof.

\bigskip
{\bf Lemma 6.}{\em If all symmetric differential polynomials $S(\p /\p z)$
in $n$ variables annihilate a polynomial $R(z)$, then $\deg R \leq \dim F_n$.}

\bigskip
{\em Proof.} The quotient of the algebra of all differential polynomials
$S(\p /\p z)$ by the ideal generated by elementary symmetric functions
is canonically isomorphic to the cohomology algebra $H^*(F_n)$. This implies
that the ideal containes the power $\frak{m} ^{\dim F_n +1}$ of the maximal
ideal $\frak{m} =(\p /\p z_1, ... , \p /\p z_n)$. This means that all
derivatives of $R$ of order $> \dim F_n$ vanish and thus $\deg R\leq \dim F_n$.

\bigskip
Proposition also implies Theorem $3$ from Introduction (describing quantum
intersection indices), since by definition of $V_n$
\[ \lan f | g \ran = [ f(\p /\p z)g(\p /\p z) V_n(z, q, c) ]|_{z=0} .\]



\begin{thebibliography}{WW}

\bibitem[A]{8} M.Atiyah, {\em Convexity and commuting hamiltonians.}
Bull. Lond. Math. Soc. {\bf 23} (1982), 1--15.

\bibitem[AB]{9} M.Atiyah, R.Bott, {\em The moment map and
equivariant cohomology.} Topology {\bf 23} (1984), 1--28.

\bibitem[CV]{3} S.Cecotti, C.Vafa, {\em Exact results for supersymmetric
sigma models.} Preprint HUTP-91/A062.

\bibitem[D]{6} B.Dubrovin, {\em Integrable systems in topological field
theory.} Nucl. Phys. {\bf B379} (1992), 627--685.

\bibitem[FF]{23} B.Feigin, E.Frenkel, {\em Integrals of motion and quantum
groups.} Preprint, 1993.

\bibitem[F1]{11} A.Floer, {\em Morse theory and lagrangian intersections.}
 J. Diff. Geom.  {\bf 28} (1988), 513--547.

\bibitem[F2]{12} A.Floer,{\em Symplectic fixed points and holomorphic spheres.}
Commun.Math.Phys. {\bf 120} (1989), 575--611.


\bibitem[G]{16} V.A.Ginsburg {\em Equivariant cohomology and Kahler geometry.}
Funct. Anal. Appl. {\bf 21:4} (1987), 271--283.

\bibitem[G1]{13} A.Givental, {\em Periodic mappings in symplectic topology.}
Funct. Anal. Appl. {\bf 23:4} (1989), 287--300.

\bibitem[G2]{14} A.Givental, {\em A symplectic fixed point theorem for toric
manifolds.} To appear in: Progress in Math., v. 93, Birhauser, Basel.

\bibitem[G3]{15} A.Givental, {\em A mirror theorem for complex projective
spaces.}, in preparation.

\bibitem[GH]{2} P.Griffits, J.Harris, {\em Principles of algebraic geometry.}
Wiley, N.Y., 1978.

\bibitem[Gr]{20} M.Gromov, {\em Pseudo-holomorphic curves in almost complex
manifolds.} Invent. Math. {\bf 82:2} (1985), 307--347.

\bibitem[HS]{18}  H.Hofer, D.Salamon, {\em Floer homology and Novikov rings.}
Preprint, 1992.

\bibitem[K]{24} M.Kontsevich, {\em $A^{\infty }$-algebras in mirror symmetry.}
Preprint, 1993.

\bibitem[O]{19} K.Ono, {\em On the Arnold conjecture for weakly monotone
symplectic manifolds.} Preprint, 1993.

\bibitem[R]{4} A.Reyman, {\em Hamiltonian systems related to graded Lie
algebras.} in: {\em Diff. Geom., Lie groups and Mechanics, III}
Zapiski Nauchn. Sem. LOMI, v. 95, Nauka, 1980 (in Russian).

\bibitem[Ru]{21} Y.Ruan, {\em Topological sigma model and Donaldson type
invariants in Gromov theory.} Preprint.

\bibitem[S]{7} V.Sadov, {\em On equivalence of Floer's and quantum cohomology.}
Preprint HUTP-93/A027.

\bibitem[V]{1} C.Vafa,{\em Topological mirrors and quantum rings.}
in: S.-T. Yau ({\em Ed.}), {\em  Essays on mirror manifolds.}
    International Press Co., Hong Kong, 1992.

\bibitem[Vt]{17} C.Viterbo, {\em The cup-product on the Thom--Smale--Witten
complex, and Floer cohomology.} To appear in: Progress in Math., v. 93,
Birkhauser, Basel.

\bibitem[W]{5} E.Witten, {\em Two-dimensional gravity and intersection theory
on moduli space.} Surveys in Diff. Geom. {\bf 1} (1991), 243--310.

\bibitem[W2]{10} E.Witten, {\em Supersymmetry and Morse theory.} J. Diff. Geom.
 17 (1982), 661--692.


\end{thebibliography}
\end{document}